\documentclass[12pt]{article}
\usepackage{amssymb,amscd,array}
\catcode `\@=11
\@addtoreset{equation}{subsection}

\def\harr#1#2{\smash{\mathop{\hbox to .5in{\rightarrowfill}}
\limits^{\scriptstyle#1}_{\scriptstyle#2}}}

\def\harrl#1#2{\smash{\mathop{\hbox to .5in{\leftarrowfill}}
\limits^{\scriptstyle#1}_{\scriptstyle#2}}}

\def\qed{\blacksquare}
\newcommand{\be}{\begin{equation}}
\newcommand{\ee}{\end{equation}}
\newcommand{\R}{\mathbb{R}}
\newcommand{\N}{\mathbb{N}}
\newcommand{\C}{\mathbb{C}}
\newtheorem{thm}{Theorem}[section]
\newtheorem{rem}[thm]{Remark}
\newtheorem{lemma}[thm]{Lemma}
\newtheorem{cor}[thm]{Corollary}
\newtheorem{prop}[thm]{Proposition}

\begin{document}
\begin{titlepage}
\begin{center}
{\bf \Large{The Standard Model and its Generalisations ~\\
in Epstein-Glaser Approach to Renormalisation Theory\\}}
\end{center}
\vskip 1.0truecm
\centerline{D. R. Grigore
\footnote{e-mail: grigore@theor1.theory.nipne.ro, grigore@theory.nipne.ro}}
\vskip5mm
\centerline{Dept. of Theor. Phys., Inst. Atomic Phys.}
\centerline{Bucharest-M\u agurele, P. O. Box MG 6, ROM\^ANIA}
\vskip 2cm
\bigskip \nopagebreak
\begin{abstract}
\noindent

\end{abstract}
\end{titlepage}

\section{Introduction}

The traditional approach to renormalisation theory starts from Bogoliubov
axioms imposed on the $S$-matrix (or equivalently on the chronological
products) and translates them into axioms on the Feynman amplitudes (the
so-called Hepp axioms). Then one tries to find explicit solutions of these
axioms using some regularization procedure and extracting in a consistent way
the ultra-violet infinities. Already for a real scalar field this task is not
very easy, but the theory becomes extremely complicated when one considers
systems with gauge invariance. Rigorous analysis performed by Becchi, Rouet and
Stora shows the tremendous complexity of the theory. In recent years, a new way
to consider renormalisation theory was advocated by prof. G. Scharf starting
from the analysis of Epstein and Glaser \cite{EG1}. In this approach one also
starts from Bogoliubov axioms on the chronological products, but tries to find
out solutions in a purely recursive way using the support properties of
various distributions appearing in the game and a procedure called distribution
splitting. More important, in this approach one can bring new light on the
problem of gauge invariance: one can argue that a consistent interaction 
(in the sense of perturbation theory) involving spin $1$ Bosons should be a
gauge invariant interaction. 

In a preceding paper \cite{Gr1} we have extended a result of Aste and Scharf
\cite{AS} concerning the uniqueness of the non-Abelian gauge theory describing
the consistent interaction of $r$ null-mass Bosons of spin $1$. We have showed
that the gauge invariance principle is a natural consequence of the description
of spin-one particles in a factor Hilbert space. One considers a auxiliary
Hilbert space 
${\cal H}^{gh}$
(describing physical and ghost particles) and assumes the existence of a
supercharge $Q$ operator acting in it. Then gauge invariance expresses the
possibility of factorising the $S$-matrix to the physical space, which is
usually constructed according to the cohomological-type formula:
$
{\cal H}_{phys} = Ker(Q)/Im(Q).
$
The obstructions to such a factorisation process are the well-known anomalies.
The main problem in this approach was the fact that this factorisation can be
implemented only in the adiabatic limit so one has to solve simultaneously the
ultra-violet and the infra-red problems. If the spin-one Bosons are massless,
then one cannot solve this combined problem. The case when the spin-one Bosons
of non-null mass are admitted in the game was studied by Aste, D\"utsch and
Scharf \cite{DS}, \cite{ASD3} for the concrete case of the electro-weak
interaction i.e. when the gauge group is exactly
$SU(2) \times U(1)$.  
In this paper we analyse the same problem considering that the spin-one Bosons
can have non-null masses and we do not impose any restriction on their number
and masses (also we do not take into account here matter fields). We will show
that in this case one can hope for a theory where the ultraviolet problems
are completely under control and the infrared problems seem to be less severe.

We should also mention that, as in \cite{Gr1}, we fill a gap in the existing
literature concerning the legitimacy of using the identification
$
{\cal H}_{phys} = Ker(Q)/Im(Q).
$
Indeed, from the physical point of view, one should proceed in strict
accordance with the dogma of the second quantization of Fock-Cook as follows:
one starts with a one-particle Hilbert space 
${\sf H}$
which is usually some representation of the Poincar\'e group; in our case we
consider a massive spin $1$ particle. Then one chooses a statistics, which in
this case should be Bose-Einstein and considers as a physical space the
associated symmetric Fock space
${\cal F}^{+}({\sf H})$.
It is not obvious that this coincides with
$
{\cal H}_{phys} = Ker(Q)/Im(Q)
$
although it is usually asserted that this follows from the analysis of Kugo and
Ojima \cite{KO}. In this paper, as in \cite{Gr1} for massless spin $1$ Bosons,
we prove that the identification of the two Hilbert spaces is a rigorous
mathematical fact. As a byproduct, we have a simpler analysis of the unitarity
of the $S$-matrix.

Summing up, the main objective of this paper is to show that the very
construction of the standard model follows from Bogoliubov axioms and the
factorization condition to the physical Hilbert space
${\cal H}_{phys}$;
these conditions determine in a rather strict way the structure of the
interaction Lagrangian for (massive) Bosons of spin $1$. To make the connection
with the Feynman graphs terminology, it will follow from the proof that only
tree diagrams of the first and second order of the perturbation theory are
considered. The remarkable fact about this approach is the fact that no
classical Lagrangian, subject afterwards to some quantization procedure, is
needed. 

We should also mention an apparent drawback of the approach, which is in fact
common to all known approaches to gauge theories. To construct the 
${\cal H}^{gh}$
one has to consider some explicit representations for the one-particle Hilbert
space (for instance to work with Wigner representation or the helicity
representation for the irreducible projective unitary representations of the
Poincar\'e group). Also one has to add a set of  ghost fields such that the the
cohomology of the supercharge gives the physical Fock space. These two choices
are non-canonical, i.e. no formul\ae~ independent of the representation are
available. Of course, it is plausible to conjecture that for two distinct
choices one gets the same physics i.e. there is a unitary transformation
intertwining them. This conjecture is suggested by the analysis of gauge
theories in the framework of classical field theory where the ghost fields are
some canonical objects associated to a principal fibre bundle.
In some particular cases, the conjecture can be actually proved. For instance,
one can consider some special choice for the Hilbert space and the ghost fields
corresponding to the so-called linear gauges. In these gauges the $S$-matrix
{\it a priori} depends on a gauge parameter and one can prove that this
dependence drops out after factorization to the physical Hilbert space 
\cite{ASD4}.

The structure of the paper is the following one. In the next Section we
generalise the description of non-null mass spin-one Bosons on similar lines as
in \cite{Gr1}. A modification of the supercharge will be necessary. As in
\cite{DS} and \cite{ASD3}, the appearance of the Higgs fields seems
unavoidable.  In Section 3 we construct the first-order $S$-matrix following
closely the lines of the computations from \cite{Gr1} to which we will
frequently refer. We will also be able to give a generic form for the
second-order $S$-matrix.  Next we impose gauge invariance for the second-order
of the perturbation theory and obtain, as expected, that the structure is
rather tight i.e. there are severe restrictions on the various coefficients of
the various Wick monomials entering the interaction Lagrangian. Moreover we
naturally obtain that some of these coefficients can be organised into a
representation of dimension $r$ of the gauge group, which is nothing else but
the representation $T$ of the Higgs fields. In particular, some very
complicated computations from \cite{DS} leading to the cancelation of possible
anomalies are nothing else but the representation property of $T$. Some
interesting mass relations connecting the structure constants, the
representation $T$ and the masses of the Bosons naturally emerge. These
relations seems to be new in the literature, at least to our knowledge, and
they have the merit of having a rigorous status. 

Finally, we test the generic formalism on the standard model of electro-weak
interactions. In this way the results of \cite{ASD3} are reobtained. In the
last Section we list some future problems which have to be solved.

We mention other papers also treating the quantisation of massive Bosons of
spin one in the Epstein-Glaser approach: \cite{D1}, \cite{D3}, 
\cite{Hu1}-\cite{Hu4}, \cite{Kr1} and \cite{Kr3}. 

Concerning the relationship between Epstein-Glaser-Scharf approach and the
traditional ways of computing various effects in the standard model (see for
instance \cite{Kraus}) we think that no serious discrepancies should appear. We
are basing this assessment on the fact that the starting point is the same:
the Bogoliubov axioms of perturbative renormalisation theory. On the other
hand, one should note that the expression of the BRST transformation in this
approach is a linearized version of the usual one. This does not mean that the
results should be different from the usual approach but this point is not
completely settled and deserves further investigations. We should also mention
a recent work of \cite{HS} on the quantum Noether method with can be probably
used to derive the same results as ours and moreover to prove the equivalence
with the results of the traditional approach.

\newpage

\section{Spin-One Relativistic Free Particles \\ 
with Positive Mass\label{mass}}
\subsection{General Description}

As in \cite{Gr1}, we take the one-particle space of the problem
${\sf H}$ 
to be the Hilbert space of a unitary irreducible representation of the
Poincar\'e group. We give below the relevant formul\ae~ for particles of mass
$m > 0$ 
and spin one. 

The upper hyperboloid of mass 
$m \geq 0$
is by definition the set of functions 
$
X^{+}_{m} \equiv \{p \in \R^{4} \vert \quad \Vert p\Vert^{2} = m^{2}\, \quad
p_{0} > 0\}
$
which are square integrable with respect to the Lorentz invariant measure
$d\alpha_{m}^{+}(p) \equiv {d{\bf p} \over 2\omega({\bf p})}$;
(in fact only classes of functions identical up to null-measure sets are
considered). The conventions are the following:
$\Vert \cdot \Vert$
is the Minkowski norm defined by
$
\Vert p \Vert^{2} \equiv p\cdot p 
$
and
$p\cdot q$
is the Minkowski bilinear form:
$
p\cdot q \equiv p_{0}q_{0} - {\bf p} \cdot {\bf q}.
$

Let us consider the 
Hilbert space
${\sf H} \equiv L^{2}(X^{+}_{m},\C^{4},d\alpha_{m}^{+})$
with the scalar product
\be
<\phi,\psi> \equiv \int_{X^{+}_{m}} d\alpha_{m}^{+}(p) \quad 
<\phi(p),\psi(p)>_{\C^{4}}
\label{scalar-prod}
\ee
where
$
<u,v>_{\C^{4}} \equiv \sum_{i=1}^{4} \overline{u_{i}} v_{i}
$
is the usual scalar product from $\C^{4}$. In this Hilbert space we have the
following (non-unitary) representation of the Poincar\'e group:
\be
\left( U_{a,\Lambda} \phi\right)(p) \equiv 
e^{ia\cdot p} \Lambda \cdot \phi(\Lambda^{-1}\cdot p)\quad {\rm for} \quad 
\Lambda \in {\cal L}^{\uparrow}, \qquad 
\left( U_{I_{t}} \phi\right)(p) \equiv \overline{ \phi(I_{s}\cdot p)}.
\label{reps0}
\ee

We define on ${\sf H}$ the following non-degenerate sesquilinear form:
\be
(\phi,\psi) \equiv = \int_{X^{+}_{m}} d\alpha_{m}^{+}(p) \quad g^{\mu\nu} 
\overline{\phi_{\mu}(p)}\psi_{\nu}(p);
\label{sesqui-form}
\ee
the indices $\mu,\nu$
take the values 
$0,1,2,3$
and the summation convention over the dummy indices is used. This form behaves
naturally with respect to the representation (\ref{reps0}).

Now we have immediately:
\begin{lemma}
Let us consider the following subspace of ${\sf H}$:
\be
{\sf H}_{m}\equiv \{ \phi \in {\sf H} \vert \quad p^{\mu} \phi_{\mu}(p) = 0\}.
\ee

Then the sesquilinear form
$\left. (\cdot,\cdot)\right|_{{\sf H}_{m}}$
is {\bf strictly} positively defined.
\label{h'}
\end{lemma}

As a consequence we have:
\begin{prop}
The representation (\ref{reps0}) of the Poincar\'e group leaves
invariant the subspace
${\sf H}_{m}$ 
and the restriction of this representation to this subspace (also denoted by 
$U$) is equivalent to the unitary irreducible representation 
${\sf H}^{[m,1]}$
of the Poincar\'e group (describing particles of mass 
$m > 0$
and spin $1$ \cite{Va}.)
\end{prop}

By definition, the couple
$({\sf H}_{m},U)$
is called a {\it spin-one Boson} of mass $m$. 

We turn now to the second quantisation procedure applied to such an elementary
system. We express the (Bosonic) {\it Fock space} of the system 
\be
{\cal F}_{m} \equiv {\cal F}^{+}({\rm H}_{m}) 
\equiv \oplus_{n\geq 0} {\cal H}'_{n}, \quad {\cal H}'_{0} \equiv \C
\label{fock}
\ee
as a subspace of an auxiliary Fock space
\be
{\cal H} \equiv {\cal F}^{+}({\rm H}) 
\equiv \oplus_{n\geq 0} {\cal H}_{n}, \quad {\cal H}_{0} \equiv \C
\label{fock-aux}
\ee

One canonically identifies the
$n^{th}$-particle subspace
${\cal H}_{n}$
with the set of Borel functions
$\Phi^{(n)}_{\mu_{1},\dots,\mu_{n}}: (X^{+}_{m})^{\times n} \rightarrow \C$
which are square integrable:
\be
\int_{(X^{+}_{m})^{\times n}} \prod_{i=1}^{n} d\alpha^{+}_{m}(k_{i})
\sum_{\mu_{1},\dots,\mu_{n} =0}^{3}
|\Phi^{(n)}_{\mu_{1},\dots,\mu_{n}}(k_{1},\dots,k_{n})|^{2} < \infty
\ee
and verify the symmetry property
\be
\Phi^{(n)}_{\mu_{P(1)},\dots,\mu_{P(n)}}(k_{P(1)},\dots,k_{P(n)}) =
\Phi^{(n)}_{\mu_{1},\dots,\mu_{n}}(k_{1},\dots,k_{n}), 
\quad \forall P \in {\cal P}_{n}.
\label{symmetry1}
\ee

In ${\cal H}$ one has natural extensions of the expression of the scalar
product (\ref{scalar-prod}) and of the sesqui-linear form (\ref{sesqui-form}).
We also have a (non-unitary) representation of the Poincar\'e group given by
the well known formula
$
{\cal U}_{g} \equiv \Gamma(U_{g}), \quad \forall g \in {\cal P};
$
here $U_{g}$ is given by (\ref{reps0}). 

Now one has from lemma \ref{h'}:
\begin{lemma}
Let us consider the following subspace of ${\cal H}$:
\be
{\cal H}' \equiv {\cal F}^{+}({\rm H}') = \oplus_{n\geq 0} {\cal H}_{n}'.
\ee

Then
${\cal H}_{n}', \quad n \geq 1$
is generated by elements of the form
$
\phi_{1} \vee \cdots \vee \phi_{n}, \quad 
\phi_{1},\dots,\phi_{n} \in {\rm H}'
$
and, in the representation adopted previously for the Hilbert space
${\cal H}_{n}$
we can take
\be
{\cal H}_{n}' = \{ \Phi^{(n)} \in {\cal H}_{n} | \quad
k_{1}^{\nu_{1}} \Phi^{(n)}_{\nu_{1},\dots,\nu_{n}}(k_{1},\dots,k_{n}) = 0\}.
\label{h-prim}
\ee

Moreover, the sesquilinear form
$\left. (\cdot,\cdot)\right|_{{\cal H}'}$
is {\bf strictly} positively defined.
\label{h'1}
\end{lemma}

Finally we have:
\begin{prop}
There exists an canonical isomorphism of Hilbert spaces
\be
{\cal F}_{m} \simeq {\cal H}'.
\ee
\label{fock-new}
\end{prop}

Now we can define the corresponding field as an operator on the Hilbert space
${\cal H}$ in complete analogy to the electromagnetic field;
we define for every
$p \in X^{+}_{m}$
the annihilation and creation operators 
\be
\left( A_{\nu}(p) \Phi\right)^{(n)}_{\mu_{1},\dots,\mu_{n}}
(k_{1},\dots,k_{n}) \equiv \sqrt{n+1}
\Phi^{(n+1)}_{\nu,\mu_{1},\dots,\mu_{n}}(p,k_{1},\dots,k_{n})
\label{annihilation-em}
\ee
and
\be
\left( A^{\dagger}_{\nu}(p) \Phi\right)^{(n)}_{\mu_{1},\dots,\mu_{n}}
(k_{1},\dots,k_{n}) \equiv - 2 \omega({\bf p}) {1\over \sqrt{n}} 
\sum_{i=1}^{n} \delta({\bf p}-{\bf k_{i}}) g_{\nu\mu_{i}} 
\Phi^{(n-1)}_{\mu_{1},\dots,\hat{\mu_{i}},\dots,\mu_{n}}
(k_{1},\dots,\hat{k_{i}},\dots,k_{n}).
\label{creation-em}
\ee

Then one has a list of properties which are formally identical with the
corresponding one from the null-mass case. First we have the {\it canonical
commutation relations} (CAR)
\be
\left[ A_{\nu}(p), A_{\rho}(p')\right] = 0,\quad
\left[ A^{\dagger}_{\nu}(p), A^{\dagger}_{\rho}(p')\right] = 0, \quad
\left[ A_{\nu}(p), A^{\dagger}_{\rho}(p')\right] = 
- 2 \omega({\bf p}) g_{\nu\rho} \delta{({\bf p}-{\bf p}')} {\bf 1}
\label{CAR-em}
\ee
and the relation
\be
(A^{\dagger}_{\nu}(p)\Psi,\Phi) = (\Psi,A_{\nu}(p)\Phi), \quad \forall 
\Psi, \Phi \in {\cal H}
\ee
which shows that
$A_{\nu}^{\dagger}(p)$
is the adjoint of 
$A_{\nu}(p)$
with respect to the sesquilinear form
$(\cdot,\cdot)$.

Next we have a natural behaviour with respect to the action of the Poincar\'e
group:
\be
{\cal U}_{a,\Lambda} A_{\nu}(p) {\cal U}^{-1}_{a,\Lambda}  = 
e^{ia\cdot p} (\Lambda^{-1})_{\nu}^{\ \rho} A_{\rho}(\Lambda \cdot p),
\quad \forall \Lambda \in {\cal L}^{\uparrow}, \quad
{\cal U}_{I_{t}} A_{\nu}(p) {\cal U}^{-1}_{I_{t}}  = 
(I_{t})_{\nu}^{\ \rho} A_{\rho}(I_{s} \cdot p)
\label{repsA-em}
\ee
and a similar relation for 
$A_{\nu}^{\dagger}(p)$.

Finally, we define the {\it field operators in the point x} according to
\be
A_{\nu}(x) \equiv A^{(+)}_{\nu}(x) + A^{(-)}_{\nu}(x)
\label{em}
\ee
where the expressions appearing in the right hand side are the positive
(negative) frequency parts and are defined by:
\be
A^{(+)}_{\nu}(x) \equiv {1\over (2\pi)^{3/2}} \int_{X^{+}_{m}}
d\alpha^{+}_{m}(p) e^{ip\cdot x} A^{\dagger}_{\nu}(p),\quad
A^{(-)}_{\nu}(x) \equiv {1\over (2\pi)^{3/2}} \int_{X^{+}_{m}}
d\alpha^{+}_{m}(p) e^{-ip\cdot x} A_{\nu}(p).
\label{em-pm}
\ee

The properties of the field operators
$A_{\nu}(x)$
are contained in the following elementary proposition:
\begin{prop}
The following relations are true:
\be
(A_{\nu}(x)\Psi,\Phi) = (\Psi,A_{\nu}(x)\Phi), \quad \forall 
\Psi, \Phi \in {\cal H},
\label{AA-dagger}
\ee
\be
(\square + m^{2}) A_{\nu}(x) = 0
\label{eqA}
\ee
and
\be
\left[A^{(\mp)}_{\mu}(x),A^{(\pm)}_{\nu}(y)\right] = 
- g_{\mu\nu} D^{(\pm)}_{m}(x-y) \times {\bf 1},\quad
\left[A^{(\pm)}_{\mu}(x),A^{(\pm)}_{\nu}(y)\right] = 0.
\label{CCR-em-pm}
\ee
As a consequence we also have:
\be
\left[A_{\mu}(x),A_{\nu}(y)\right] = - g_{\mu\nu} D_{m}(x-y) \times {\bf 1};
\label{commutation-em}
\ee
here
\be
D_{m}(x) = D_{m}^{(+}(x) + D_{m}^{(-)}(x)
\ee
is the Pauli-Jordan distribution and
$D^{(\pm)}_{m}(x)$
are given by:
\be
D_{m}^{(\pm)}(x) \equiv \pm {1\over (2\pi)^{3/2}} \int_{X^{+}_{m}}
d\alpha^{+}_{m}(p) e^{\mp i p\cdot x}.
\label{D0}
\ee
\end{prop}

Let us note that we have:
\be
(\square + m^{2}) D_{m}^{(\pm)}(x) = 0, \quad (\square + m^{2}) D_{m}(x) = 0.
\ee

We turn now on the constructions of observables on the Fock space of the
spin-one Boson
${\cal F}_{m} \simeq {\cal H}'$
by self-adjoint operators on the Hilbert space
${\cal H}$. 
If $O$ is such an operator on 
${\cal H}'$
then it induces naturally an operator (also denoted by $O$) on
${\cal H}$
which leaves invariant the subspace
${\cal H}'$. 
This type of observables on
${\cal H}$
are called {\it gauge invariant observables}. 

The description of possible interactions between the spin-one field and 
matter follows the same ideas. Let us consider that the (Fock) space of
the ``matter" fields is denoted by 
${\cal H}_{matter}$. 
Then, in the hypothesis of weak coupling, one can argue that the Hilbert space
of the combined system is
$ 
{\cal H}_{total} \equiv {\cal F}_{m} \otimes {\cal H}_{matter}.  
$ 
It is easy to see that, if we define, 
$
\tilde{\cal H} \equiv {\cal H} \otimes {\cal H}_{matter}, \quad 
$
and
$
\tilde{\cal H}' \equiv {\cal H}' \otimes {\cal H}_{matter}
$
we have as before:
\be
{\cal H}_{total} \simeq \tilde{\cal H}'.
\ee

In the Hilbert space
$\tilde{\cal H}$
we can define as usual the expressions for the spin-one field and all
properties listed previously stay true. Typical interactions terms have the 
form
\be
T_{1}(x) \equiv A_{\nu}(x) j^{\nu}(x)
\label{inter}
\ee
where
$j^{\nu}(x)$
are some Wick polynomials in the ``matter" fields called {\it currents}.  Then
conservation of the current is a sufficient and necessary condition such that
the expression (\ref{inter}) induces, in the adiabatic limit, a well defined
expression on the Hilbert space
${\cal H}_{total}$.

For higher-order chronological products, one can establish a similar
expression: 
\be
T_{n}(x_{1},\dots,x_{n}) \equiv \sum_{k=0}^{n} 
:A_{\nu_{1}}(x_{1})\cdots A_{\nu_{k}}(x_{k}): \quad
j^{\nu_{1},\dots,\nu_{k}}(x_{1},\dots,x_{n})
\label{inter-n}
\ee
and the condition of factorization, in the adiabatic limit, amounts again to
the conservation of the multi-currents
$j^{\nu_{1},\dots,\nu_{k}}(x_{1},\dots,x_{n})$.
The conservation of these multi-currents can be heuristically connected with
the gauge invariance of the $S$-matrix (see \cite{Sc1} ch. 4.6).

\newpage

\subsection{Quantisation with Ghost Fields\label{gh}}

In this subsection we give an alternative description of the Fock space
${\cal F}_{m}$
using the ghosts fields following rather closely the arguments from \cite{Gr1}.
However, in the case of positive mass particles it seems that it is not
sufficient to introduce the Fermionic ghosts and one has also to introduce a
Bosonic ghost.

In \cite{Gr1}, the Hilbert space was constructed by acting on the vacuum state
with the electromagnetic potentials
$A_{\nu}$
and the pair of ghost fields of null mass and Fermi statistics
$u$ and $\tilde{u}$.
In the case of spin-one Bosons of mass $m > 0$, we generate
${\cal H}^{gh}$
by acting on the vacuum with the potentials
$A_{\nu}$
and the triplet of ghost fields of the same mass 
$u,~\tilde{u}$
and
$\Phi$
such that the first two are Fermionic and the last one is a Bosonic field.
We will need some explicit representation for
${\cal H}^{gh}$;
taking into account the general structure outlined above, we should have:
\be
{\cal H}^{gh} = \oplus_{n,w,l,s=0}^{\infty} {\cal H}_{nwls}
\ee
where one can identify
$
{\cal H}_{nwls}
$
with the set of Borel functions
$
\Phi^{(nwls)}_{\mu_{1},\dots,\mu_{n}}: (X^{+}_{0})^{n+w+l+s} \rightarrow \C
$
which are square integrable with respect to the product measure
$(\alpha_{m}^{+})^{\times (n+w+l+s)}$:
\be
\sum_{n,w,l,s=0}^{\infty} \int_{(X^{+}_{m})^{n+w+l+s}} d\alpha^{+}_{m}(K)
d\alpha^{+}_{m}(P) d\alpha^{+}_{m}(Q) d\alpha^{+}_{m}(R) 
\sum_{\mu_{1},\dots,\mu_{n}=0}^{3} 
|\Phi^{(nwls)}_{\mu_{1},\dots,\mu_{n}}(K;P;Q;R)| \leq \infty
\ee
(here 
$
K \equiv (k_{1},\dots,k_{n}), \quad
P \equiv (p_{1},\dots,p_{w}), \quad
Q \equiv (q_{1},\dots,q_{l})
$
and
$
R \equiv (r_{1},\dots,r_{l}))
$
and verify the symmetry property
\begin{eqnarray}
\Phi^{(nwls)}_{\mu_{P(1)},\dots,\mu_{P(n)}}
(k_{P(1)},\dots,k_{P(n)};p_{Q(1)},\dots,p_{Q(w)};q_{R(1)},\dots,q_{R(l)};
r_{T(1)},\dots,r_{T(s)})
\nonumber \\
= (-1)^{|Q|+|R|} \Phi^{(nwls)}_{\mu_{1},\dots,\mu_{n}}
(k_{1},\dots,k_{n};p_{1},\dots,p_{w};q_{1},\dots,q_{l};r_{1},\dots,r_{s}), 
\nonumber \\
\forall P \in {\cal P}_{n}, Q \in {\cal P}_{w}, R \in {\cal P}_{l}, 
T \in {\cal P}_{s}.
\label{sym-nml}
\end{eqnarray}

In this representation we can construct the following annihilation operators 
\begin{eqnarray}
\left( A_{\nu}(t) \Phi\right)^{(nwls)}_{\mu_{1},\dots,\mu_{n}}
(k_{1},\dots,k_{n};P;Q;R) = 
\Phi^{(n+1,wls)}_{\nu,\mu_{1},\dots,\mu_{n}}
(t,k_{1},\dots,k_{n};P;Q;R)
\end{eqnarray}
\begin{eqnarray}
\left( b(t) \Phi\right)^{(nwls)}_{\mu_{1},\dots,\mu_{n}}
(K;p_{1},\dots,p_{w};Q;R) = 
\Phi^{(n,w+1,ls)}_{\mu_{1},\dots,\mu_{n}}
(K;t,p_{1},\dots,p_{w};Q;R)
\end{eqnarray}
\begin{eqnarray}
\left( c(t) \Phi\right)^{(nwls)}_{\mu_{1},\dots,\mu_{n}}
(K;P;q_{1},\dots,q_{l};R) = 
(-1)^{w} \sqrt{l+1}
\Phi^{(nw,l+1,s)}_{\mu_{1},\dots,\mu_{n}}
(K;P;t,q_{1},\dots,q_{l};R)
\end{eqnarray}
and
\begin{eqnarray}
\left( a(t) \Phi\right)^{(nwls)}_{\mu_{1},\dots,\mu_{n}}
(K;P;Q;r_{1},\dots,r_{s}) = 
\sqrt{s+1}
\Phi^{(nwl,s+1)}_{\mu_{1},\dots,\mu_{n}}
(K;P;Q;t,r_{1},\dots,r_{s});
\end{eqnarray}
similar expressions can be written for the creation operators.  They verify
usual canonical (anti)commutation relations and behave naturally with respect
to Poincar\'e transform.

Then the fields
\be
u(x) \equiv {1\over (2\pi)^{3/2}} \int_{X^{+}_{m}} d\alpha^{+}_{m}(q)
\left[ e^{-i q\cdot x} b(q) + e^{i q\cdot x} c^{*}(q) \right],
\label{u}
\ee
\be
\tilde{u}(x) \equiv {1\over (2\pi)^{3/2}} \int_{X^{+}_{m}} d\alpha^{+}_{m}(q)
\left[ - e^{-i q\cdot x} c(q) + e^{i q\cdot x} b^{*}(q) \right],
\label{u-tilde}
\ee
and
\be
\Phi(x) \equiv {1\over (2\pi)^{3/2}} \int_{X^{+}_{m}} d\alpha^{+}_{m}(q)
\left[ e^{-i q\cdot x} a(q) + e^{i q\cdot x} a^{*}(q) \right]
\label{phi}
\ee
are called {\it Fermionic} (resp. {\it Bosonic) ghost fields}.

They verify the wave equations:
\be
(\square + m^{2}) u(x) = 0, \quad (\square + m^{2}) \tilde{u}(x) = 0, \quad
(\square + m^{2}) \Phi(x) = 0
\label{equ}
\ee
and we have usual canonical (anti)commutation relations:
\be
\{u(x),\tilde{u}(y)\} = D_{m}(x-y) {\bf 1}, \quad
[\Phi(x), \Phi(y) ] = D_{m}(x-y) {\bf 1}
\ee
and all other (anti)commutators are null. Now we can define the operator:
\be
Q \equiv \int_{X^{+}_{m}} d\alpha^{+}_{m}(q) \left[ k^{\mu}
\left( A_{\mu}(k) c^{*}(k) + A^{\dagger}_{\mu}(k) b(k)\right) +
im \left( a(k) c^{*}(k) - a^{*}(k) b(k)\right) \right]
\label{supercharge}
\ee
called {\it supercharge}. Its properties are summarised in the following
proposition which can be proved by elementary computations:
\begin{prop}
The following relations are valid:
\be
Q \Phi_{0} = 0;
\label{Q-0}
\ee
\begin{eqnarray}
\left[ Q, A^{\dagger}_{\mu}(k) \right] = - k_{\mu} c^{*}(k),\quad
\left\{ Q, b^{*}(k) \right\} = k^{\mu} A^{\dagger}_{\mu}(k) - im a^{*}(k), 
\nonumber \\
\left\{ Q, c^{*}(k) \right\} = 0, \quad
\left[ Q, a^{*}(k) \right] = im c^{*}(k);
\label{com-dagger}
\end{eqnarray}
\begin{eqnarray}
\left[ Q, A_{\mu}(k) \right] = k_{\mu} b(k),\quad
\left\{ Q, b(k) \right\} = 0,
\nonumber \\
\left\{ Q, c(k) \right\} = k^{\mu} A_{\mu}(k) + im a(k), \quad
\left[ Q, a(k) \right] = im b(k);
\label{com}
\end{eqnarray}
\begin{eqnarray}
\{Q,u(x)\} = 0, \quad 
\{Q,\tilde{u}(x)\} =  - i \left( \partial^{\mu} A_{\mu}(x) + m \Phi(x)\right),
\nonumber \\
\left[ Q, A_{\mu}(x) \right] = i \partial_{\mu} u(x), \quad
\left[ Q, \Phi(x)\right] = im u(x);
\label{Q-com}
\end{eqnarray}
\be
Q^{2} = 0;
\label{square}
\ee
\be
Im(Q) \subset Ker(Q)
\label{im-ker}
\ee
and 
\be
{\cal U}_{g} Q = Q {\cal U}_{g}, \quad \forall g \in {\cal P}.
\label{UQ}
\ee

Moreover, one can express the supercharge in terms of the ghosts fields as
follows:
\be
Q = \int_{\R^{3}} d^{3}x  \left( \partial^{\mu} A_{\mu}(x) + m \Phi(x) \right)
\stackrel{\leftrightarrow}{\partial_{0}}u(x).
\ee
\end{prop}

(The succession of the preceding formul\ae~ suggests the most convenient way to
derive them; for instance, from (\ref{com}) and (\ref{com-dagger}) one derives
that 
$\{Q,Q\} = 0$ 
and gets (\ref{square})).  In particular (\ref{square}) justifies the
terminology of supercharge and (\ref{im-ker}) indicates that it might be
interesting to take the quotient.  Indeed, we will rigorously prove that this
quotient coincides with 
${\cal F}_{m}$.

We can give the explicit expression of the supercharge in this representation;
starting from the definition (\ref{supercharge}) we immediately get:
\begin{eqnarray}
\left(Q \Phi\right)^{(nwls)}_{\mu_{1},\dots,\mu_{n}}(K;P;Q;R) =
\nonumber \\
(-1)^{w} \sqrt{n+1\over l} \sum_{i=1}^{l} (-1)^{i-1} q^{\nu}_{i}
\Phi^{(n+1,w,l-1,s)}_{\nu,\mu_{1},\dots,\mu_{n}}
(q_{i},K;P;q_{1},\dots,\hat{q_{i}},\dots,q_{l};R)
\nonumber \\
- \sqrt{w+1\over n} \sum_{i=1}^{n} (k_{i})_{\mu_{i}}
\Phi^{(n-1,w+1,l,s)}_{\mu_{1},\dots,\hat{\mu_{i}},\dots,\mu_{n}}
(k_{1},\dots,\hat{k_{i}},\dots,k_{n};k_{i},P;Q;R)
\nonumber \\
+ im (-1)^{w} \sqrt{s+1\over l} \sum_{i=1}^{l} (-1)^{i-1} 
\Phi^{(n,w,l-1,s+1)}_{\mu_{1},\dots,\mu_{n}}
(K;P;q_{1},\dots,\hat{q_{i}},\dots,q_{l};q_{i},R)
\nonumber \\
- im \sqrt{w+1\over s} \sum_{i=1}^{s} 
\Phi^{(n,w+1,l,s-1)}_{\mu_{1},\dots,\mu_{n}}
(K;r_{i},P;Q;r_{1},\dots,\hat{r_{i}},\dots,r_{s})
\label{Q-explicit}
\end{eqnarray}
where, of course, we use Bourbaki convention 
$\sum_{\emptyset} \equiv 0$.

Now we introduce on 
${\cal H}^{gh}$
a {\it Krein operator} according to:
\be
\left( J\Phi\right)^{(nwls)}(K;P;Q;R) \equiv 
(-1)^{wl} (-g)^{\otimes n} \Phi^{(nlws)}(K;Q;P;R).
\label{Krein-gh}
\ee

The properties of this operator are contained in the following proposition:
\begin{prop}
The following relations are verified:
\be
J^{*} = J^{-1} = J
\ee
\be
J b(p) J = c(p), \quad J c(p) J = b(p), \quad 
J A^{*}_{\mu}(p) J = A^{\dagger}_{\mu}(p), \quad J a(p) J = a(p)
\ee
\be
J Q J = Q^{*}
\ee
and
\be
{\cal U}_{g} J = J {\cal U}_{g}, \quad \forall g \in {\cal P}.
\label{UJ}
\ee
Here $O^{*}$ is the adjoint of the operator $O$ with respect to the scalar
product 
$<\cdot,\cdot>$
on
${\cal H}^{gh}$.
\end{prop}

We define, as usual, a sesquilinear form on 
${\cal H}^{gh}$ according to
\be
(\Psi,\Phi) \equiv <\Psi,J\Phi>;
\label{sesqui-gh}
\ee
then this form is non-degenerate.  It is convenient to denote the conjugate of
the arbitrary operator $O$ with respect to the sesquilinear form
$(\cdot,\cdot)$
by
$O^{\dagger}$ 
i.e.
\be
(O^{\dagger} \Psi,\Phi) = (\Psi,O \Phi).
\ee

Then the following formula is available:
\be
O^{\dagger} = J O^{*} J.
\ee

As a consequence, we have
\be
A_{\mu}(x)^{\dagger} = A_{\mu}(x), \quad
u(x)^{\dagger} = u(x), \quad
\tilde{u}(x)^{\dagger} = - \tilde{u}(x), \quad
\Phi(x)^{\dagger} = \Phi(x).
\label{conjugate}
\ee

From (\ref{UJ}) it follows that we have:
\be
({\cal U}_{g}\Psi,{\cal U}_{g}\Phi) = (\Psi,\Phi), 
\forall g \in {\cal P}^{\uparrow}, \quad
({\cal U}_{I_{t}}\Psi,{\cal U}_{I_{t}}\Phi) = \overline{(\Psi,\Phi)}.
\label{inv-sesqui}
\ee

As in \cite{Gr1}, we give a description of the factor space 
$Ker(Q)/Im(Q)$. 
We will construct a ``homotopy" for the supercharge $Q$.

\begin{prop}
Let us define the operator
\be
\tilde{Q} \equiv - {1\over 2m^{2}} \int_{X^{+}_{m}} d\alpha^{+}_{m}(q) \left[
k^{\mu}  \left( A_{\mu}(k) b^{*}(k) + A^{\dagger}_{\mu}(k) c(k)\right) +
im \left(a^{*}(k) c(k) - a(k) b^{*}(k)\right) \right].
\label{tildeQ}
\ee

Then the following relation is valid:
\be
Y \equiv \{Q,\tilde{Q}\} = N_{a} + N_{b} + N_{c} + X
\label{Y}
\ee
where 
$N_{a}$
($N_{c}, N_{c}$)
are particle number operators for the ghosts of type $a$ (resp. $b, c$) and
\be
X \equiv - {1\over m^{2}} \int_{X^{+}_{m}} d\alpha^{+}_{m}(k) k^{\mu} k^{\nu}
A^{\dagger}_{\mu}(k) A_{\nu}(k).
\label{X}
\ee

Moreover the following relations are true:
\be
\tilde{Q}^{2} = 0
\ee
and
\be
[Y, Q] = 0, \quad [Y, \tilde{Q}] = 0.
\label{YQ}
\ee
\end{prop}

The operator 
$\tilde{Q}$
it is called the {\it homotopy} of $Q$. The operator $Y$ is not invertible, 
but as in \cite{Gr1} we have:
\begin{prop}
The operator
$\left. Y\right|_{{\cal H}_{nwls}}$ is invertible iff 
$w + l +s > 0$.
\end{prop}

{\bf Proof:}
An alternative expression for the operator $X$ defined by (\ref{X}) is:
\be
X = A \otimes {\bf 1}
\ee 
where the operator $A$ acts only on the Bosonic variables and is given by the
expression
\be
A = d\Gamma(P);
\ee
here
$d\Gamma$
is the familiar Cook functor defined by;
\be
d\Gamma(P) \psi_{1} \otimes \cdots \otimes \psi_{n} \equiv
P\psi_{1} \otimes \psi_{2} \cdots \otimes \psi_{n} + \cdots 
\psi_{1} \otimes \psi_{2} \cdots \otimes P\psi_{n} 
\ee
and the operator $P$ is in our case given by:
\be
(P \psi)_{\mu}(k) \equiv {1\over m^{2}} k_{\mu}k^{\nu} \psi^{\nu}(k).
\ee

We immediately obtain that $P$ is a projector i.e.
$P^{2} = P$
and we have, as in the case of massless Bosons of spin $1$, the direct sum
decomposition of the one-particle Bosonic subspace into the direct sum of
$Ran(P)$ 
and 
$Ran(1-P)$.  
Let us consider a basis in the one-particle Bosonic subspace formed by a basis
$f_{i}, \quad i \in \N$ 
of 
$Ran(P)$ 
and a basis 
$g_{i}, \quad i \in \N$ 
of
$Ran(1-P)$.
A basis in the $n^{\rm th}$-particle Bosonic subspace is of the form:
$$
f_{i_{1}} \vee \cdots f_{i_{r}} \vee g_{j_{1}} \vee \cdots \vee g_{j_{t}},
\quad r, t \in \N, \quad r + t = n.
$$

Applying the operator $A$ to such a vector gives the same vector multiplied by
$r$. So, in the basis chosen above, the operator $A$ is diagonal with diagonal
elements from $\N$. It follows that the operator 
$\left. Y\right|_{{\cal H}_{nwls}}$ 
can also be exhibited into a diagonal form with diagonal elements of the form
$w + l + s + r , \quad r \in \N$.
It is obvious that for
$w + l +s > 0$
this is an invertible operator.
$\qed$

Accordingly, we have the following corollary:
\begin{cor}
Let us define
$
{\cal H}_{0} \equiv \oplus_{n \geq 0} {\cal H}_{n000}
$
and
$
{\cal H}_{1} \equiv \oplus_{n \geq 0, w+l+s > 0} {\cal H}_{nwls}
$.
Then the operator $Y$ has the block-diagonal form
\begin{eqnarray}
Y = \left( \matrix{ Y_{1} & 0 \cr 0 & Y_{0} } \right) 
\end{eqnarray}
with 
$Y_{1}$
an invertible operator.
\end{cor}
and the the fundamental result
\begin{prop}
There exists the following vector spaces isomorphism:
\be
Ker(Q)/Im(Q) \simeq {\cal H}'
\label{factor}
\ee
where the subspace
${\cal H}'$
has been defined in the previous subsection (see the lemmas \ref{h'1})
\label{cohomology}
\end{prop}

{\bf Proof:}
(i)  As in \cite{Gr1} one can prove that if
$
\Phi \in Ker(Q)
$
then we have the decomposition
\be
\Phi = Q \psi + \tilde{\Phi}
\ee
where
\be
\tilde{\Phi}^{(nwls)} = 0, \quad w + l + s > 0.
\ee
The condition 
$Q\Phi = 0$
amounts now to
$Q\tilde{\Phi} = 0$
or, with the explicit expression of the supercharge (\ref{Q-explicit}):
\be
q^{\nu} \tilde{\Phi}^{(n+1,0,0,0)}_{\nu,\mu_{1},\dots,\mu_{n}}(q,k_{1},\dots,
k_{n};\emptyset;\emptyset;\emptyset) = 0,
\quad \forall n \in \N
\ee
i.e. the ensemble
$\left. \{\tilde{\Phi}^{(n00)}\}\right|_{n \in \N}$
is an element from
${\cal H}'$
(see lemma \ref{h'1}).

It remains to see in what conditions such 
$\tilde{\Phi}$
is an element from 
$Im(Q)$
i.e. we have
$\tilde{\Phi} = Q \chi$.
It is clear that only the components
$\chi^{(n100)}$
should be taken non-null. Then the expression of the supercharge
(\ref{Q-explicit}) gives the following condition:
\begin{eqnarray}
\left(Q \chi\right)^{(n001)}_{\mu_{1},\dots,\mu_{n}} (K;\emptyset;\emptyset;r) 
= - im  \chi^{(n100)}_{\mu_{1},\dots,\mu_{n}}(K;r;\emptyset;\emptyset) = 0.
\label{Q-chi}
\end{eqnarray}

Because the mass $m$ of the Boson is non-null, we get
$
\chi = 0 \quad \Rightarrow \tilde{\Phi} = 0
$
and we obtain the assertion from the statement.
$\qed$

We finally get as in \cite{Gr1}:

\begin{thm}
The isomorphism (\ref{factor}) extends to a Hilbert space isomorphism:
$$
\overline{Ker(Q)/Im(Q)} \simeq {\cal F}_{m}
$$
and the factorised representation of the Poincar\'e group coincides with the
representation acting into the space
${\cal H}'.$
\label{photon+ghosts}
\end{thm}

We close with an important observation. One can easily see that one can take 
the limit
$m \searrow 0$
in the expressions for the various Hilbert spaces and quantum fields and also
on the expression of the supercharge $Q$. (The expression $\tilde{Q}$ does not
have the limit in the obvious way, but this is not very important, because this
expression had played only an auxiliary r\^ole). In this limit we can write
\be
{\cal H}^{gh} \simeq {\cal H}^{gh}_{0} \otimes {\cal H}_{\Phi}
\ee
where
${\cal H}^{gh}_{0}$
is the  Hilbert space generated by the fields
$A_{\mu}(x),~u(x),~\tilde{u}(x)$
and 
${\cal H}_{\Phi}$
is generated by the scalar ghosts. Then the supercharge (\ref{supercharge})
takes the form
\be
Q = Q' \otimes {\bf 1}
\ee
where $Q'$ coincides formally with the expression of $Q$ for
$m \searrow 0$
but acts only in
${\cal H}^{gh}_{0}$.
Moreover, we have:
\be
\overline{Ker(Q)/Im(Q)} \simeq \overline{Ker(Q')/Im(Q')} 
\otimes {\cal H}_{\Phi}
\ee
i.e. we can see that the states from
${\cal H}_{\Phi}$
decouple completely and can be considered {\bf physical}. Moreover, one can
see that, in this case, nothing prevents us to consider that the scalar
``ghost" has a non-zero mass. This observation is essential for the
construction of the standard model, because a scalar ``ghost" field
corresponding to a null mass Boson, if considered a physical field of non-zero
mass is nothing else but the {\it Higgs field} \cite{ASD3}. Let us stress that
this observation has to be raised at the status of a postulate; it agrees with
the construction of the standard model as we shall see in the following and
brings no mathematical inconsistencies.

\newpage
\subsection{Gauge-Invariant Observables\label{gau-obs}}

As in \cite{Gr1}, we denote by ${\cal W}$ the linear space of all Wick 
monomials on the Fock space
${\cal H}^{gh}$
i.e. containing the fields 
$A_{\mu}(x),~ u(x),~\tilde{u}(x)$
and
$\Phi(x)$.
If $M$ is such a Wick monomial, we define by
$gh_{\pm}(M)$
the degree in 
$\tilde{u}$ (resp. in $u$). 
The {\it ghost number} is, by definition, the expression:
\be
gh(M) \equiv gh_{+}(M) - gh_{-}(M)
\ee
i.e. we conserve the same expression as in the massless case. The {\it BRST 
operator} also has the same expression: it is given by
\be
d_{Q} M \equiv :QM: - (-1)^{gh(M)} :MQ:
\label{BRST-op}
\ee
on monomials $M$ and extend it by linearity to the whole 
${\cal W}$. 

Most of the formul\ae~from \cite{Gr1} stay true:
\be
d_{Q}^{2} = 0,
\label{Q2}
\ee
\be
d_{Q} u = 0, \quad 
d_{Q} \tilde{u} = - i (\partial^{\mu} A_{\mu} + m \Phi), \quad
d_{Q} A_{\mu} = i \partial_{\mu} u, \quad
d_{Q} \Phi = i m u;
\label{BRST}
\ee
\be
d_{Q}(MN) = (d_{Q}M) N + (-1)^{gh(M)} M (d_{Q}N), \quad
\forall M, N \in {\cal W}.
\label{Leibnitz}
\ee

The class of {\it all} observables on the factor space emerges (see theorem 
\ref{photon+ghosts}): an operator
$O: {\cal H}^{gh} \rightarrow {\cal H}^{gh}$
induces a well defined operator 
$[O]$
on the factor space
$\overline{Ker(Q)/Im(Q)} \simeq {\cal F}_{m}$
if and only if it verifies:
\be
\left. d_{Q} O \right|_{Ker(Q)} = 0.
\label{dQ}
\ee

Not all operators verifying the condition (\ref{dQ}) are interesting. In fact,
the operators of the type
$d_{Q} O$
are inducing a null operator on the factor space; explicitly, we have:
\be
[d_{Q} O] = 0.
\ee

Moreover, in this case the following formula is true for the matrix elements of
the factorized operator 
$[O]$:
\be
([\Psi], [O] [\Phi]) = (\Psi, O \Phi). 
\label{matrix-elem}
\ee

If the interaction Lagrangian is a Wick monomial
$T_{1} \in {\cal W}$
with
$gh(T_{1}) \not= 0$
then the $S$-matrix is trivial.

The analysis of the possible interactions between the Bosonic spin-one field
and ``matter" follows the usual lines (see \cite{Gr1}). Let
${\cal H}_{matter}$
be the corresponding Hilbert space of the matter fields; it is elementary to 
see that we can realise the total Hilbert space
${\cal H}_{total} \equiv {\cal F}_{m} \otimes {\cal H}_{matter}$
as the factor space 
$Ker(Q)/Im(Q)$
where the supercharge $Q$ is defined on
$\tilde{\cal H}_{gh} \equiv {\cal H}_{gh} \otimes {\cal H}_{matter}$
by the obvious substitution
$Q \rightarrow Q \otimes {\bf 1}$.

We define on
$\tilde{\cal H}_{gh}$
the interaction Lagrangian of the same form (\ref{inter}) where the current
$j^{\mu}(x)$
is a Wick polynomial and it is conserved.

\newpage
\section{Massive Yang-Mills Fields\label{ym}}

\subsection{The General Setting}

As in \cite{Gr1}, we first define in an unambiguous way what we mean by
Yang-Mills fields. The main modification is that now all the fields will carry
an additional index
$a = 1,\dots,r$
and this can be realised with an appropriate modification of the Hilbert spaces
(auxiliary or physical). So we have the fields:
$
A_{a\mu},~ u_{a},~\tilde{u}_{a},~\Phi_{a} \quad a = 1,\dots,r
$
given by the following expressions:
\be
A_{a\mu}(x) \equiv {1\over (2\pi)^{3/2}} 
\int_{X^{+}_{m_{a}}} d\alpha^{+}_{m_{a}}(p) 
\left[ e^{-ip\cdot x} A_{a\mu}(p) + e^{ip\cdot x} A^{\dagger}_{a\mu}(p)\right]
\label{YM}
\ee
\be
u_{a}(x) \equiv {1\over (2\pi)^{3/2}} 
\int_{X^{+}_{m_{a}}} d\alpha^{+}_{m_{a}}(q)
\left[ e^{-i q\cdot x} b_{a}(q) + e^{i q\cdot x} c_{a}^{\dagger}(q) \right]
\label{u-YM}
\ee
\be
\tilde{u}_{a}(x) \equiv {1\over (2\pi)^{3/2}} 
\int_{X^{+}_{m_{a}}} d\alpha^{+}_{m_{a}}(q)
\left[ - e^{-i q\cdot x} c_{a}(q) + e^{i q\cdot x} b_{a}^{\dagger}(q) \right]
\label{u-tilde-YM}
\ee
and
\be
\Phi_{a}(x) \equiv {1\over (2\pi)^{3/2}} 
\int_{X^{+}_{m_{a}}} d\alpha^{+}_{m_{a}}(q)
\left[ e^{-i q\cdot x} a_{a}(q) + e^{i q\cdot x} a_{a}^{\dagger}(q) \right].
\label{phi-YM}
\ee

As in \cite{Gr1}, this amounts to consider that the one-particle subspace is a
direct sum of $r$ copies of elementary heavy Bosons of masses 
$m_{a}, \quad a = 1,\dots,r$ 
and spin $1$.

These fields verify the following equations of motion:
\be
(\square + m_{a}^{2}) u_{a}(x) = 0, \quad 
(\square + m_{a}^{2}) \tilde{u}_{a}(x) = 0, \quad
(\square + m_{a}^{2}) \Phi_{a}(x) = 0, \quad a = 1,\dots,r.
\label{equ-r}
\ee

The canonical (anti)commutation relations are:
\begin{eqnarray}
\left[A_{a\mu}(x),A_{b\nu}(y)\right] = - 
\delta_{ab} g_{\mu\nu} D_{m_{a}}(x-y) \times {\bf 1},
\nonumber \\
\{u_{a}(x),\tilde{u}_{b}(y)\} = \delta_{ab} D_{m_{a}}(x-y) \times {\bf 1}, 
\quad
[ \Phi_{a}(x),\Phi_{b}(y) ] = \delta_{ab} D_{m_{a}}(x-y) \times {\bf 1}
\label{comm-r}
\end{eqnarray}
and all other (anti)commutators are null.  The supercharge is given by (see
(\ref{supercharge})):
\be
Q \equiv \sum_{a=1}^{r} 
\int_{X^{+}_{m_{a}}} d\alpha^{+}_{m_{a}}(q) \left[ k^{\mu} 
\left( A_{a\mu}(k) c_{a}^{*}(k) + A^{\dagger}_{a\mu}(k) b_{a}(k)\right)
 + i m_{a} 
\left( a_{a}(k) c_{a}^{\dagger}(k) - a^{*}_{a}(k) b_{a}(k)\right) \right]
\label{supercharge-YM}
\ee
and verifies all the expected properties. 

The Krein operator has an expression similar to (\ref{Krein-gh}) and can be 
used to construct a sesquilinear form like in (\ref{sesqui-gh}). Then relations
of the type (\ref{conjugate}) are still true;
\be
A_{a\mu}(x)^{\dagger} = A_{a\mu}(x), \quad
u_{a}(x)^{\dagger} = u_{a}(x), \quad
\tilde{u}_{a}(x)^{\dagger} = - \tilde{u}_{a}(x), \quad
\Phi_{a}(x)^{\dagger} = \Phi_{a}(x).
\label{conjugate-YM}
\ee

As a consequence, proposition \ref{cohomology}, and the main theorem
\ref{photon+ghosts} stay true.

The ghost degree is defined in an obvious way and the expression of the BRST
operator (\ref{BRST-op}) is the same in this more general framework and the
corresponding properties are easy to obtain. In particular we have (see
(\ref{BRST})):
\be
d_{Q} u_{a} = 0, \quad 
d_{Q} \tilde{u}_{a} = - i (\partial_{\mu} A_{a}^{\mu} + m_{a} \Phi_{a}) , \quad
d_{Q} A_{a}^{\mu} = i \partial^{\mu} u_{a}, \quad
d_{Q} \Phi_{a} = i m_{a} u_{a}, \quad \forall a = 1,\dots,r.
\label{BRST-YM}
\ee

If we take into account the last observation from the preceding Subsection, it
appears that it is possible to make in the formalism presented above some of
the masses null. In this case the corresponding scalar ghosts can be considered
as physical fields and they will be called {\it Higgs fields}. Moreover, we do
not have to assume that they are massless i.e. if some Boson field
$A_{a}^{\mu}$
has zero mass
$m_{a} = 0$,
we can suppose that the corresponding Higgs field
$\Phi_{a}$
has a non-zero mass:
$m^{H}_{a}$.
If we use the compact notation 
\be
m^{*}_{a} \equiv \left\{\begin{array}{rcl} 
m_{a} & \mbox{for} & m_{a} \not= 0 \\
m^{H}_{a} & \mbox{for} & m_{a} = 0\end{array}\right. 
\ee
this will imply that the last equation (\ref{equ-r}) will be replaced by
\be
[\square + (m^{*}_{a})^{2}] \Phi_{a}(x) = 0
\ee
and the last relation of (\ref{comm-r}) will become:
\be
[ \Phi_{a}(x),\Phi_{b}(y) ] = \delta_{ab} D_{m^{*}_{a}}(x-y) \times {\bf 1};
\ee
here
$a = 1,\dots,r$.

Moreover, this process of attributing a non-zero mass to the scalar partners of
the zero-mass vector fields should not influence the BRST transformation
formula (\ref{BRST-YM}); that is, this formula remains unchanged.  We raise
this comment to the status of postulate as at the end of the preceding
Subsection and we make some comments. Suppose that we have $s$ zero-mass
spin-one Bosons and 
$r-s$ 
non-null spin-one Bosons in out theory; here
$s \leq r$ 
is arbitrary. The consistency of the formalism requires that the scalar
partners of everyone of the non-null spin-one Bosons should be a fictious
particle - a ghost - with exactly the same mass. On the contrary, the scalar
partners of the $s$ zero-mass spin-one Bosons can be taken as {\it physical}
particles of arbitrary mass. This postulate is in agreement with the case of
the standard model, where we have exactly one zero-mass Boson - the photon -
and exactly one physical scalar partner - the Higgs particle. Moreover, this
postulate do not bring mathematical inconsistencies. However, we should mention
that this postulate has its limitations: for instance it is possible that in
the case when all Bosons are massive there are no solutions in this framework.
In other words, models like the Higgs-Kibble model (see \cite {KO}) are not
covered by this postulate. Recently \cite{Sc2}, professor Scharf noticed that
one can consider a consistent theory with 
$t > s$
physical scalars and in this framework Higgs-Kibble type models can be
described, although their physical relevance is not so clear for the moment.
(See also ref. \cite{DS1}).

We will construct a perturbation theory {\it \'a la} Epstein-Glaser for the
{\it free} fields
$A_{a}^{\mu},~ u_{a},~\tilde{u}_{a}$
and
$\Phi_{a}, \quad a = 1,\dots,r$
in the auxiliary Hilbert space
${\cal H}_{YM}^{gh,r}$
imposing the usual axioms of causality, unitarity and relativistic invariance.
Moreover, we want that the result factorizes to the physical Hilbert space in
the adiabatic limit. This amounts to

\be
lim_{\epsilon \searrow 0} 
\left. d_{Q} \int_{(\R^{4})^{\times n}} dx_{1} \cdots dx_{n}
g_{\epsilon}(x_{1}) \cdots g_{\epsilon}(x_{n}) 
T_{n}(x_{1},\dots,x_{n}) \right|_{Ker(Q)} = 0, 
\quad \forall n \geq 1.
\label{gi-Tn}
\ee

If this condition if fulfilled, then the chronological and the
antichronological products do factorize to the physical Hilbert space and they
give a perturbation theory verifying causality, unitarity and relativistic
invariance.

We have to comment on this point with is rather delicate. One can argue that
there are reasons to believe that the infra-red (or adiabatic) limit do not
exists so the preceding relation does not have a rigorous status. Moreover, in
the adiabatic limit tri-linear Wick monomials are zero and this affects the
argument of the next theorem. This seems to jeopardize the very nice physical
interpretation of this relation, namely as a consistency condition (the
factorization to the physical space of the $S$-matrix) which avoids the
necessity to consider gauge invariance as a separate postulate. This problem is
avoided by the Z\"urich group as follows \cite{DHKS1} - \cite{DHS3}. One admits
instead of (\ref{gi-Tn}) an ``infinitesimal" version, namely:
\be
d_{Q} T_{n}(x_{1},\dots,x_{n}) = i \sum_{l=1}^{n} 
{\partial \over \partial x^{\mu}_{l}} T^{\mu}_{n/l}(x_{1},\dots,x_{n})
\label{zurich}
\ee
where
$T^{\mu}_{n/l}(x_{1},\dots,x_{n})$
are some Wick polynomials which should be determined recurringly, together
with the chronological products. This implies that, instead of (\ref{gi-Tn}) we
have
\be
\left. d_{Q} \int_{(\R^{4})^{\times n}} dx_{1} \cdots dx_{n}
g_{\epsilon}(x_{1}) \cdots g_{\epsilon}(x_{n}) 
T_{n}(x_{1},\dots,x_{n}) \right|_{Ker(Q)} = O(\epsilon) 
\quad \forall n \geq 1
\label{gi-epsilon}
\ee
i.e. the factorization is valid only up to terms of order $\epsilon$.

One should not touch the adiabatic limit, that is one should construct the
matrix
$S(g)$
for
$g \not\equiv 1$.
We can mention here that the basis for this new postulate is in fact the
relation (\ref{gi-Tn}) considered as a heuristic idea. So the two relations
are rather closely related and there is no severe opposition between them.
But in this way, one will construct an object
$S(g)$
which {\bf does not factorize to the physical space}
$Ker(Q)/Im(Q)$
as can be easily seen. This raises doubts about its physical interpretation.

Another way out would be to restrict the physical space 
${\cal F}_{m}$
even further, that is to consider that only some of the states of this space
are physically accessible, for instance only those states containing ``soft
photons", like in the usual treatments of the infra-red divergences. Then one
would have to modify the factorization condition (\ref{gi-Tn}) (replacing
$Ker(Q)$
by a smaller subspace) and check that this does not invalidate some of the
arguments from the next Subsection, like the linear independence arguments.

So, an honest point of view is the following one: the adiabatic problem is
still open and it is presumably the main obstacle left for the construction of
a complete rigorous version of the standard model. We conjecture that a nice
``cure" of this problem can be found and in that case one we will be able to
accept the consistency condition (\ref{gi-Tn}) as rigorous mathematical fact.
In this case gauge invariance will not be an independent postulate, as we do
advocate here. Till then, if we want to be completely rigorous, we are forced
to replace (\ref{gi-Tn}) by (\ref{zurich}) and deal afterwards with the
adiabatic limit in the usual and rather unsatisfactory ways.

Let us close this Subsection with an important remark. One sees that the
expressions (\ref{BRST-YM}) are only the linearized part of the usual BRST
transformation. This has the important consequence that (\ref{BRST-YM}) is
compatible with the equations of motion (this can be seen by applying the
Klein-Gordon operator on both sides of the relations). In other words, the
transformation (\ref{BRST-YM}) is well defined {\it on shell}. On the contrary,
the full BRST transformation is defined only {\it off shell}, i.e. is not
compatible with the equations of motion (because of the non-linear terms) so it
cannot be unitary implemented in the Hilbert space. This implies that in the
usual approaches to Yang-Mills theories, one must construct from {\it on shell}
objects (like the chronological products) some {\it off shell} objects (like
the generating function of the Green functions) on which one can acts with the
full (non-linear) BRST transformation. Then one must prove that the full BRST
transform is equivalent to the factorization condition of the $S$-matrix
(\ref{gi-epsilon}).

\newpage

\subsection{The Derivation of the Yang-Mills Lagrangian; \\
First-Order Gauge Invariance}

In this subsection we completely exploit the condition of gauge invariance in
the first order of perturbation theory obtaining the generic form of the
Yang-Mills interaction of spin-one Bosons. We assume the summation convention
of the dummy indices $a,b,\dots.$

We define the {\it canonical dimension} 
$\omega(W)$
of a Wick monomial $W$ by atributting the value $1$ (resp. $3/2$) to every
integer spin field factor or derivative (resp.  half-integer spin field factor)
and summing up the values of all factors. For a Wick polynomial, the canonical
dimension is the supremum of the canonical dimensions of all summands.

\begin{thm}
Let us consider the operator
\be
T_{1}(g) = \int_{\R^{4}} dx~ g(x) T_{1}(x)
\ee
defined on 
${\cal H}^{gh,r}_{YM}$
with
$T_{1}$
a Lorentz-invariant Wick polynomial in 
$A_{\mu},~u,~\tilde{u}$
and
$\Phi$
verifying also
$\omega(T_{1}) \leq 4$.
If
$T_{1}(g)$
induces a well defined non-trivial $S$-matrix, in the adiabatic limit, then
it necessarily has the following form: 
\be
T_{1}(g) = \int_{\R^{4}} dx~ g(x) \left[ T_{11}(x) + T_{12}(x) + T_{13}(x) + 
T_{14}(x) + T_{15}(x) + T_{16}(x) \right]
\label{YM-1}
\ee
where we have introduced the following notations:
\be
T_{11}(x) \equiv
f_{abc} \left[ :A_{a\mu}(x)A_{b\nu}(x) \partial^{\nu} A_{a}^{\mu}(x): -
:A_{a}^{\mu}(x) u_{b}(x) \partial_{\mu} \tilde{u}_{c}(x):\right],
\ee
\be
T_{12}(x) \equiv 
f'_{abc} \left[ :\Phi_{a}(x) \partial_{\mu} \Phi_{b}(x) A_{c}^{\mu}(x): 
- m_{b} :\Phi_{a}(x) A_{b\mu}(x) A_{c}^{\mu}(x): 
- m_{b} :\Phi_{a}(x) \tilde{u}_{b}(x) u_{c}(x):\right]
\ee
\be
T_{13}(x) \equiv 
f^{"}_{abc} :\Phi_{a}(x) \Phi_{b}(x) \Phi_{c}(x): 
\ee
\be
T_{14}(x) \equiv 
g_{abcd} :\Phi_{a}(x) \Phi_{b}(x) \Phi_{c}(x) \Phi_{d}(x): 
\ee
\be
T_{15} \equiv h_{ab} \left[ : A_{a\mu}(x)A_{b}^{\mu}(x): -
2 :\tilde{u}_{a}(x) u_{b}(x):\right], \quad
T_{16}(x) \equiv 
h'_{ab} :\Phi_{a}(x) \Phi_{b}(x): .
\ee

Here the various constants from the preceding expression are constrained by the
following conditions:

- the expressions 
$f_{abc}$
are completely antisymmetric
\be
f_{abc} = - f_{bac} = - f_{acb}
\label{anti-f}
\ee
and verify:
\be
(m_{a} - m_{b}) f_{abc} = 0, \quad {\rm iff} \quad m_{c} = 0, \quad 
\forall a,b = 1,\dots,r;
\label{mass-f}
\ee

- the expressions
$f'_{abc}$
are antisymmetric  in the indices $a$ and $b$:
\be
f'_{abc} = - f'_{bac},
\label{anti-f'}
\ee
verify the relation:
\be
(m^{H}_{a} - m^{H}_{b}) f'_{abc} = 0, \quad 
{\rm iff} \quad m_{a} = m_{b} = m_{c} = 0,
\quad \forall a,b = 1,\dots,r
\label{mass-f'}
\ee
and are connected to 
$f_{abc}$
by:
\be
f_{abc} m_{c} = f'_{cab} m_{a} - f'_{cba} m_{b}, \quad 
\forall a,b,c = 1,\dots,r;
\label{f-f'}
\ee

- the expressions 
$f^{"}_{abc}$
are completely symmetric in all indices and remain undetermined for
$m_{a} = m_{b} = m_{c} = 0;$
for the opposite case they are given by:
\be
f^{"}_{abc} = {1 \over 6m_{c}} f'_{abc} 
\left[(m^{*}_{a})^{2} - (m^{*}_{b})^{2} - m_{a}^{2} + m_{b}^{2}\right],
\label{f"}
\ee
for
$m_{c} \not=0$. 

- the expressions 
$g_{abcd}$
are non-zero if and only if
$
m_{a} = m_{b} = m_{c} = m_{d} = 0
$
and they are completely symmetric;

- the expressions
$h_{ab}$
are symmetric
\be
h_{ab} = h_{ba}
\ee
and verify the relation
\be
(m_{a} - m_{b}) h_{ab} = 0,  \quad \forall a,b = 1,\dots,r;
\label{mass-h}
\ee

- the constants
$h'_{ab}$
are undetermined for
$m_{a} = m_{b} = 0$
and in the opposite case are given by:
\be
h'_{ab} = {m_{a} \over 2 m_{b}} h_{ab}, \quad {\rm iff} \quad m_{b} \not=0, 
\label{h}
\quad \forall a = 1,\dots,r.
\ee
(We note that it is implicit in relations like (\ref{mass-f}), (\ref{mass-f'}),
etc. that the summation convention over the dummy indices does not apply).
\label{T1}
\end{thm}

{\bf Proof:}
(i) We follow closely the line of argument of theorem 4.1 from \cite{Gr1}.
If we take into account Lorentz invariance, the power counting condition
from the statement and the restriction of non-triviality 
$gh(T_{1}) = 0$
the list of linearly independent Wick monomials from \cite{Gr1} (formula 4.2.4
from Subsection 4.2) is enlarged by new possibilities containing, of course, 
the scalar ghosts:

$\bullet$ of degree 2:
\be
T^{(1)'} = h^{(1)}_{ab} :A_{a\mu}(x)~ A_{b}^{\mu}(x): \quad
T^{(2)'} = h^{(2)}_{ab} :\tilde{u}_{a}(x)~ u_{b}(x): \quad
T^{(3)'} = h^{(3)}_{ab} :\Phi_{a}(x)~ \Phi_{b}(x):
\label{list2}
\ee
$\bullet$ of degree 3: 
\begin{eqnarray}
T^{(1)"} = h^{(1)}_{abc} :\Phi_{a}(x)~ A_{b\mu}(x)~ A_{c}^{\mu}(x):, \quad
T^{(2)"} = h^{(2)}_{abc} :\Phi_{a}(x)~ \tilde{u}_{b}(x)~ u_{c}(x):
\nonumber \\
T^{(3)"} = h^{(3)}_{abc} :\Phi_{a}(x)~ \Phi_{b}(x)~ \Phi_{c}(x):, \quad
T^{(4)"} = h^{(4)}_{ab} :\Phi_{a}(x)~ \partial_{\mu} A_{b}^{\mu}(x):, \quad
\nonumber \\
T^{(5)"} = h^{(5)}_{ab} :\partial_{\mu} \Phi_{a}(x)~ A_{b}^{\mu}(x):
\label{list3}
\end{eqnarray}

$\bullet$ of degree 4:
\begin{eqnarray}
T^{(1)} = f^{(1)}_{abc} :A_{a\mu}(x)~ A_{b\nu}(x)~
\partial^{\nu} A_{c}^{\mu}(x): \quad
T^{(2)} = f^{(2)}_{abc} :A_{a}^{\mu}(x)~ u_{b}(x)
\partial_{\mu} \tilde{u}_{c}(x): \quad
\nonumber \\
T^{(3)} = f^{(3)}_{abc} :A_{a}^{\mu}(x)~ \partial_{\mu} u_{b}(x)~
\tilde{u}_{c}(x): \quad
T^{(4)} = f^{(4)}_{abc} :\partial_{\mu} A_{a}^{\mu}(x)~ u_{b}(x)
\tilde{u}_{c}(x): \quad
\nonumber \\
T^{(5)} = f^{(5)}_{abc} :A_{a\mu}(x)~ A_{b}^{\mu}(x)~
\partial_{\nu} A_{c}^{\nu}(x): \quad
T^{(6)} = g^{(1)}_{abcd} :A_{a\mu}(x)~ A_{b}^{\mu}(x)~
A_{c\nu}(x)~ A_{d}^{\nu}(x):
\nonumber \\
T^{(7)} = g^{(2)}_{abcd} :A_{a\mu}(x)~ A_{b}^{\mu}(x)~
u_{c}(x)~ \tilde{u}_{d}(x): \quad
T^{(8)} = g^{(3)}_{abcd} :u_{a}(x)~ u_{b}(x)~
\tilde{u}_{c}(x)~ \tilde{u}_{d}(x):
\nonumber \\
T^{(9)} = g^{(4)}_{abcd} \varepsilon_{\mu\nu\rho\sigma}
:A_{a}^{\mu}(x)~ A_{b}^{\nu}(x)~ A_{c}^{\rho}(x)~ A_{d}^{\sigma}(x): \quad
T^{(10)} = g^{(1)}_{ab} :\partial_{\mu} A_{a\nu}(x)~
\partial^{\mu} A_{b}^{\nu}(x):
\nonumber \\
T^{(11)} = g^{(2)}_{ab} :\partial_{\mu} A_{a}^{\mu}(x)~
\partial_{\nu} A_{b}^{\nu}(x): \quad
T^{(12)} = g^{(3)}_{ab} :\partial_{\mu} A_{a\nu}(x)~
\partial^{\nu} A_{b}^{\mu}(x):
\nonumber \\
T^{(13)} = g^{(4)}_{ab} :A_{a}^{\mu}(x)~
\partial_{\mu} \partial_{\nu} A_{b}^{\nu}(x): \quad
T^{(14)} = g^{(5)}_{ab} \varepsilon_{\mu\nu\rho\sigma}
:F_{a}^{\mu\nu}(x)~ F_{b}^{\rho\sigma}(x):
\nonumber \\
T^{(15)} = g^{(6)}_{ab} :\partial_{\mu} u_{a}(x)~
\partial^{\mu} \tilde{u}_{b}(x):
\nonumber \\
T^{(16)} = f^{(6)}_{abc} :\Phi_{a}(x)~ \Phi_{b}(x)~ 
\partial_{\mu} A_{c}^{\mu}(x): \quad 
T^{(17)} = f^{(7)}_{abc} :\Phi_{a}(x)~ \partial_{\mu} \Phi_{b}(x) 
A_{c}^{\mu}(x): \quad 
\nonumber \\
T^{(18)} = g^{(5)}_{abcd} :\Phi_{a}(x)~ \Phi_{b}(x)~ 
A_{c\mu}(x)~ A_{d}^{\mu}(x): \quad
T^{(19)} = g^{(6)}_{abcd} :\Phi_{a}(x)~ \Phi_{b}(x)~ 
\tilde{u}_{c}(x)~ u_{d}(x): 
\nonumber \\
T^{(20)} = g^{(7)}_{abcd} :\Phi_{a}(x)~ \Phi_{b}(x)~ \Phi_{c}(x)~ \Phi_{d}(x):
\quad
T^{(21)} = h^{(5)}_{ab} :\partial_{\mu}\Phi_{a}(x)~ \partial^{\mu}\Phi_{b}(x):
\quad
\label{list4}
\end{eqnarray}

Without losing generality we can impose the following symmetry restrictions on
the constants from the preceding list:
\begin{eqnarray}
h^{(1)}_{ab} = h^{(1)}_{ba} \quad
g^{(1)}_{abcd} = g^{(1)}_{bacd} = g^{(1)}_{abdc} = g^{(1)}_{cdab} \quad
g^{(2)}_{abcd} = g^{(2)}_{bacd}
\nonumber \\
h^{(3)}_{ab} = h^{(3)}_{ba}, \quad
h^{(5)}_{ab} = h^{(5)}_{ba}, \quad
h^{(1)}_{abc} = h^{(1)}_{acb}, \quad 
\nonumber \\
g^{(3)}_{abcd} = - g^{(3)}_{bacd} = - g^{(3)}_{abdc}  \quad
g^{(i)}_{ab} = g^{(2)}_{ba} \quad i = 1, 2, 3, 5
\nonumber \\
g^{(5)}_{abcd} = g^{(5)}_{bacd} = g^{(5)}_{abdc}, \quad
g^{(6)}_{abcd} = g^{(6)}_{bacd} 
\label{s1}
\end{eqnarray}
and one can suppose that
$g^{(4)}_{abcd}$ (resp. $ g^{(7)}_{abcd}$)
are completely antisymmetric (resp. symmetric) in all indices.

(ii) By integration over $x$ some of the linear independence is lost in the
adiabatic limit. (In the language of axiom (\ref{zurich}), some of the terms
can be grouped in total divergences.) Namely, all the conclusions from
\cite{Gr1} stay true and we have in the end:
\begin{itemize}
\item
One can eliminate
$T^{(3)}$
by redefining the constants
$f^{(2)}_{abc}$
and
$f^{(4)}_{abc}$;
\item
One can eliminate
$T^{(5)}$
by redefining the constants
$f^{(1)}_{abc}$;
\item
One can eliminate
$T^{(12)}$
and
$T^{(13)}$
by redefining the constants
$g^{(2)}_{ab}$;
\item
One can eliminate
$T^{(10)}$
and
$T^{(15)}$
using the equation of motion (\ref{eqA}) and (\ref{equ});
\item
$T^{(14)}$
is null in the adiabatic limit.
\item
One can eliminate
$T^{(5)"}$
by redefining the constants
$h^{(4)}_{ab}$;
\item
One can choose the constants
$f^{(7)}_{abc}$
such that they verify
\be
f^{(7)}_{abc} = - f^{(7)}_{bac}
\ee
if one modifies 
$f^{(6)}_{abc}$
appropriately.
\item
One can eliminate 
$T^{(21)}$
by redefining the constants
$h^{(3)}_{ab}$.
\end{itemize}

(iii) Some of the remaining expressions are of the form
$d_{Q} O$
so they do not count. Namely
\begin{itemize}
\item
We have
$$
d_{Q} : \left(\partial_{\mu} A_{a}^{\mu} + m_{a} \Phi_{a} \right)\tilde{u}_{b}:
 = : \left(\partial_{\mu} A_{a}^{\mu} + m_{a} \Phi_{a}\right)
\left(\partial_{\mu} A_{b}^{\mu} + m_{b} \Phi_{b} \right):
$$
so we can give up the expressions
$T^{(11)}$
if we modify appropriately the expressions
$h^{(3)}_{ab}$
and
$f^{(6)}_{abc}$
conveniently; afterwards one can trade off
$f^{(6)}_{abc}$
modifying
$f^{(7)}_{abc}$
by integration by parts as explained above;
\item
We have
$$
d_{Q} : \tilde{u}_{a} \Phi_{b}: = 
-i : \left(\partial_{\mu} A_{a}^{\mu} + m_{a} \Phi_{a}\right) \Phi_{b}: +
im_{b} :\tilde{u}_{a} u_{b}:
$$
so we can eliminate the expression
$T^{(4)"}$
if we redefine the expressions
$h^{(3)}_{ab}$
and
$h^{(2)}_{ab}$;
\item
If the constants
$g_{abc}$
are chosen antisymmetric in the indices $a$ and $c$, then we have
$$
d_{Q} g_{abc} :\tilde{u}_{a} u_{b} \tilde{u}_{c}: = 
2i g_{abc} : \left( \partial_{\mu} A^{\mu}_{a} + m_{a} \Phi_{a} \right)
u_{b} \tilde{u}_{c}:
$$
so, it follows that if we modify conveniently the constants
$h^{(2)}_{abc}$
we can impose
\be
f^{(4)}_{abc} = f^{(4)}_{cba};
\label{s2}
\ee
\item
We have
$$
d_{Q} : \Phi_{a} \Phi_{b} \tilde{u}_{c}: =
im_{a} :u_{a} \Phi_{b} \tilde{u}_{c}: 
+ im_{b} :\Phi_{a} u_{b} \tilde{u}_{c}: 
- :\Phi_{a} \Phi_{b} (\partial_{\mu} A^{\mu}_{c} + m_{c} \Phi_{c}): 
$$
so we can give up the term
$T^{(16)}$
if we modify conveniently the constants
$h^{(2)}_{abc}$
and
$h^{(3)}_{abc}$.
\end{itemize}

(iv) As a conclusion, we can keep in
$T_{1}$
only the expressions
$
T^{(1)'} - T^{(3)'},~ T^{(1)"} - T^{(3)"},~T^{(1)}, \\ T^{(2)},~ T^{(4)}~,
T^{(6)}-T^{(9)}
$
and
$T^{(17)}-T^{(20)}$
with the appropriate symmetry properties.

We compute now the expression
$d_{Q} T_{1}$;
the expression (4.2.7) from \cite{Gr1} gets new contributions:
\begin{eqnarray}
d_{Q} T_{1} = d_{Q} T_{1}^{(0)} 
-i h^{(2)}_{ab} m_{a} :\Phi_{a} u_{b}: 
+ i f^{(2)}_{abc} m_{c} :A_{a}^{\mu} u_{b} \partial_{\mu} \Phi_{c}: 
-i f^{(4)}_{abc} m_{c} :\partial_{\mu} A_{b}^{\mu} u_{b} \Phi_{c}: 
\nonumber \\
-i g^{(2)}_{abcd} m_{d} :A_{a\mu} A_{b}^{\mu} u_{c} \Phi_{d}: 
- 2i g^{(3)}_{abcd} m_{d} :u_{a} u_{b} \tilde{u}_{c} \Phi_{d}: 
+ i f^{(1)}_{abc} m_{c}^{2} :A_{a\mu} u_{b} A_{c}^{\mu}:
\nonumber \\ 
-i f^{(4)}_{abc} m_{a}^{2} :u_{a} u_{b} \tilde{u}_{c}: 
+2i h^{(3)}_{ab} m_{a} :u_{a} \Phi_{b}: 
\nonumber \\
+ i f^{(7)}_{abc} \left[m_{a} :u_{a} \partial_{\mu} \Phi_{b} A_{c}^{\mu}: 
+ m_{b} :\Phi_{a} \partial_{\mu}u_{b} A_{c}^{\mu}: + 
:\Phi_{a} \partial^{\mu} \Phi_{b} \partial_{\mu} u_{c}: \right]
\nonumber \\
+ i h^{(1)}_{abc} \left[ m_{a} :u_{a} A_{b\mu} A_{c}^{\mu}: 
+ 2 :\Phi_{a} A_{b}^{\mu} \partial_{\mu} u_{c}: \right]
+ i h^{(2)}_{abc} \left[ m_{a} :u_{a} \tilde{u}_{b} u_{c}: 
-:\Phi_{a} \left(\partial_{\mu}A_{b}^{\mu} + m_{b}\Phi_{b}\right) u_{c}:\right]
\nonumber \\
+ 3i h^{(3)}_{abc} m_{a} :u_{a} \Phi_{b} \Phi_{c}: 
+ 2 i g^{(5)}_{abcd} \left[ m_{a} :u_{a} \Phi_{b} A_{c\mu} A_{d}^{\mu}: 
+ :\Phi_{a} \Phi_{b} A_{c}^{\mu} \partial_{\mu}u_{d}: \right]
\nonumber \\
+ i g^{(6)}_{abcd} \left[ 2m_{a} :u_{a} \Phi_{b} \tilde{u}_{c} u_{d}: 
- :\Phi_{a} \Phi_{b} \left( \partial_{\mu}A_{c}^{\mu} + m_{c} \Phi_{c} \right) 
u_{d}: \right]
+ 4i g^{(7)}_{abcd} m_{a} :u_{a} \Phi_{b} \Phi_{c} \Phi_{d}: \qquad
\end{eqnarray}

Here 
\begin{eqnarray}
d_{Q} T_{1}^{(0)} = i \partial_{\mu} [
2 h^{(1)}_{ab} :A_{a}^{\mu} u_{b}: +
( f^{(1)}_{abc} - f^{(1)}_{cba}) :u_{a} A_{b\nu} \partial^{\nu} A^{\mu}_{c}:
+ f^{(1)}_{bac} :u_{a} A_{b\nu} \partial^{\mu} A^{\nu}_{c}:
\nonumber \\
+ f^{(1)}_{bca} :\partial_{\nu}u_{a} A_{b}^{\nu} A^{\mu}_{c}: -
f^{(1)}_{cba} :u_{a} \partial^{\nu}A_{b\nu} A_{c\nu}: ]
\nonumber \\
- i (2 h^{(1)}_{ab} + h^{(2)}_{ab} ) :\partial_{\mu}A_{a}^{\mu} u_{b}:
+i ( f^{(1)}_{cba} - f^{(1)}_{abc})
:u_{a} \partial_{\mu}A_{b\nu} \partial^{\nu}A^{\mu}_{c}:
\nonumber \\
+ i ( f^{(1)}_{abc} - f^{(1)}_{bac} - f^{(1)}_{cba} + f^{(2)}_{cba})
:\partial_{\mu}\partial_{\nu}A_{a}^{\mu} A_{b}^{\nu} u_{c}: +
i (f^{(1)}_{abc} + f^{(4)}_{acb})
:\partial_{\mu}A_{a}^{\mu} \partial_{\nu}A_{b}^{\nu} u_{c}:
\nonumber \\
- i f^{(1)}_{acb}:\partial_{\nu}A_{a\mu} \partial^{\nu}A_{b}^{\mu} u_{c}: +
i f^{(2)}_{abc}:\partial^{\mu}u_{a} u_{b} \partial_{\mu}\tilde{u}_{c}:
+ 4i g^{(1)}_{abcd} :\partial_{\mu}u_{a} A_{b}^{\mu} A_{c\nu} A_{d}^{\nu}:
\nonumber \\
+ i g^{(2)}_{abcd} ( 2 :\partial_{\mu}u_{a} A_{b}^{\mu} u_{c} \tilde{u}_{d}: +
:A_{a\mu} A_{b}^{\mu} u_{c} \partial_{\rho} A^{\rho}_{d}: )
- 2 i g^{(3)}_{abcd} :u_{a} u_{b} \partial_{\mu}A_{c}^{\mu} \tilde{u}_{d}:
\nonumber \\
-4 i g^{(4)}_{abcd} \varepsilon_{\mu\nu\rho\sigma}
:\partial^{\mu}u_{a} A_{b}^{\nu} A_{c}^{\rho} A_{d}^{\sigma}:
\label{zero}
\end{eqnarray}
is the expression (4.2.7) from \cite{Gr1}, i.e. the expression
$d_{Q} T_{1}$
for zero-mass Bosons and without scalar ghosts, the next terms having various
origins: the modification of the BRST transformation (\ref{BRST-YM}), the
modification of the equation of motion (\ref{equ-r}) and the new terms
$T^{(3)'},\quad T^{(1)"} - T^{(3)"}$
and
$T^{(17)}-T^{(20)}$
considered in the expression of
$T_{1}$.
We impose the condition of factorisation to the physical space 
(\ref{gi-epsilon}) for the case
$n = 1$:
\be
\left. \int_{\R^{4}} dx~ g_{\epsilon}(x) d_{Q} T_{1}(x)\right|_{Ker(Q)} = 
O(\epsilon).
\label{gi-1}
\ee

It is not very hard to see that all the conclusions from \cite{Gr1} remain 
true, i.e. the constants
$f_{abc} \equiv f^{(1)}_{abc}$
are completely antisymmetric,
\be
f^{(2)}_{abc} = - f_{abc}, \quad f^{(4)}_{abc} = 0, \quad
g^{(i)}_{abcd} = 0, \quad i = 1,2,3,4, \quad
2h^{(1)}_{ab} + h^{(2)}_{ab} = 0.
\ee
Moreover, we get: 
\be
2 h^{(1)}_{abc} m_{a} = f_{bac} (m_{b}^{2} - m_{c}^{2}), 
\quad \forall a,b,c = 1,\dots,r,
\label{1}
\ee
\be
h_{ab} m_{a} = 2 h^{(3)}_{ab} m_{b}, \quad \forall a,b = 1,\dots,r,
\label{2}
\ee
\be
g^{(i)}_{abcd} = 0, \quad i = 5,6;
\ee
and the expressions
$
g^{(7)}_{abcd} 
$
can be non-zero {\it iff}
$
m_{a} = m_{b} = m_{c} = m_{d} = 0.
$ 

It remains to perform some integrations by parts into the remaining expression
and to obtain:
\begin{eqnarray}
d_{Q} T_{1} = i \partial_{\mu} \left [\cdots 
+ \left( 2 h^{(1)}_{cab} + f^{(7)}_{cba} m_{b}\right) 
:A_{a}^{\mu} u_{b} \Phi_{c}: 
+ f^{(7)}_{abc} :\Phi_{a} \partial^{\mu} \Phi_{b} u_{c}: \right]
\nonumber \\
- i (f_{abc} m_{c}^{2} + h^{(2)}_{acb} m_{a}) :u_{a} u_{b} \tilde{u}_{c}: 
+i (- f_{abc} m_{c} + 2 f^{(7)}_{bca} m_{b} - 2 h^{(1)}_{cab})
:A_{a}^{\mu} u_{b} \partial_{\mu}\Phi_{c}: 
\nonumber \\
- i (2 h^{(1)}_{acb} + f^{(7)}_{abc} m_{b} + h^{(2)}_{acb})
:\Phi_{a} u_{b}\partial_{\mu}A_{c}^{\mu}: -
i \left[h^{(2)}_{abc} m_{b} - 3 h^{(3)}_{abc} m_{c} -
f^{(7)}_{abc} (m^{H}_{b})^{2} \right] :\Phi_{a} \Phi_{b} u_{c}: \quad
\end{eqnarray}
where by 
$\cdots$
we mean the expression obtained if all the masses are zero and there are no
scalar ghosts (see (\ref{zero}) above.)  The divergence contributes in 
(\ref{gi-1}) with a term of order 
$\epsilon$
and the other terms can be computed on vectors from
${\cal H}'$.
In this way we see that we get independent conditions from each term in the
preceding formula i.e.
\be
2 f_{abc} m_{c}^{2} = h^{(2)}_{bca} m_{b} -  h^{(2)}_{acb} m_{a},
\quad \forall a,b,c = 1,\dots, r,
\label{3}
\ee
\be
2 h^{(1)}_{cab} = - f_{abc} m_{c} + 2 f^{(7)}_{bca} m_{b},
\quad \forall a,b,c = 1,\dots, r,
\label{4}
\ee
\be
h^{(2)}_{abc} = - 2 h^{(1)}_{abc} - f^{(7)}_{acb} m_{c},
\quad \forall a,b,c = 1,\dots, r,
\label{5}
\ee
and
\be
6 h^{(3)}_{abc} m_{c} = 
f^{(7)}_{abc} \left[ (m^{H}_{a})^{2} - (m^{H}_{b})^{2} \right] +
h^{(2)}_{bac} m_{a} + h^{(2)}_{abc} m_{b},
\quad \forall a,b,c = 1,\dots, r.
\label{6}
\ee

We exploit completely the system of equations (\ref{1}), (\ref{3}) - (\ref{6}).
It is obvious that in order to obtain the statement of the theorem we should
redefine
$f^{(7)}_{abc} \rightarrow f'_{abc}$,
$h^{(3)}_{abc} \rightarrow f^{"}_{abc}$
and
$h^{(3)}_{ab} \rightarrow h'_{ab}$.
If we take the symmetric (resp. antisymmetric) part in $a$ and $b$ of the
relation (\ref{4}) we get an explicit expression for $h^{(1)}_{cab}$:
\be
h^{(1)}_{cab} = {1\over 2} ( f'_{bca} m_{b} + f'_{acb} m_{a}),
\quad \forall a,b,c = 1,\dots, r
\label{h1}
\ee
and respectively the consistency relation (\ref{f-f'}).  One substitutes this
result into the equations (\ref{1}) (resp. (\ref{5})) and gets an identity
(resp. an explicit expression for
$h^{(2)}_{abc}$):
\be
h^{(2)}_{abc} = f'_{abc} m_{b}, \quad \forall a,b,c = 1,\dots, r.
\label{h2}
\ee

Next, from (\ref{6}) for
$m_{c} = 0$
we get the consistency relation (\ref{mass-f'}) and for
$m_{c} \not= 0$
we obtain the expression (\ref{f"}).

Finally, from (\ref{2}) we immediately get the consistency relation
(\ref{mass-h}) and the explicit expression (\ref{h}). If the expressions for
$h^{(1)}_{abc}$
and
$h^{(2)}_{abc}$
are substituted into the generic expression for
$T_{1}$
we get the formula from the statement.

(vi) It remains to prove that the expression from the statement cannot be of
the type
$d_{Q} O$
and this can be easily done.
$\qed$

\begin{rem}
It is a remarkable fact that we get in a natural way mass relations of the type
(\ref{mass-f}). This relations is non-trivial iff there are simultaneously
massive and massless Bosons in the model. In this case, we can reformulate this
relation as follows: if 
$f_{abc} \not= 0$
and
$m_{c} = 0$
then necessarily we have
$m_{a} = m_{b}$. In particular this is the cause of the equality of the masses
of the two heavy $W$ Bosons in the standard model. 
\end{rem}

The relation (\ref{f-f'}) can be completely exploited:
\begin{cor}
The following relations are true:
\be
f'_{abc} = {m_{a}^{2} + m_{b}^{2} - m_{c}^{2} \over 2 m_{a} m_{b}} f_{abc},
\quad \forall a,b,c = 1,\dots,r \quad {\rm s. t.} 
\quad m_{a} \not= 0, \quad m_{b} \not= 0
\label{f'1}
\ee
and
\be
f'_{abc} = m_{c} g_{abc}, \quad \forall a,b,c = 1,\dots,r \quad {\rm s. t.} 
\quad m_{a} = 0, \quad m_{b} \not= 0.
\label{f'2}
\ee

Here the constants
$g_{abc}$
are constrained only by the symmetry property in the last two indices
\be
g_{abc} = g_{acb}.
\label{sym-g}
\ee

These two relations are completely equivalent to the relation (\ref{f-f'}) 
so, in particular the constants
$f'_{abc}$
remain arbitrary for
$m_{a} = m_{b} = 0.$
\end{cor}

{\bf Proof:}
The first relation can be obtained if we multiply (\ref{f-f'}) by 
$m_{c}$
and perform two cyclic permutations. Combining the three relations in a
convenient way one gets (\ref{f'1}). The relation (\ref{f'2}) follows from
(\ref{f-f'}) if we consider the case
$m_{c} = 0.$
$\qed$

\begin{cor}
In the condition of the preceding theorem, one has:
\be
d_{Q} T_{1}(x) = i\partial_{\mu} T_{1}^{\mu}(x)
\label{k1}
\ee
where:
\be
T_{1}^{\mu} \equiv T_{11}^{\mu} + T_{12}^{\mu} + T_{13}^{\mu}
\ee
and the expression from this formula are defined as follows:
\be
T_{11}^{\mu}  \equiv f_{abc} \left( :u_{a} A_{b\nu} F^{\nu\mu}_{c}: -
{1\over 2} :u_{a} u_{b} \partial^{\mu} \tilde{u}_{c}: \right),
\ee
\be
T_{12}^{\mu} \equiv f'_{abc} \left( m_{a} :A_{a}^{\mu} \Phi_{b} u_{c}:
+ :\Phi_{a} \partial^{\mu}\Phi_{b} u_{c}: \right),
\ee
and
\be
T_{13}^{\mu} \equiv  2 h_{ab} :A_{a}^{\mu} u_{b}:. 
\ee
\end{cor}

Moreover, we have:
\begin{prop}
The expression
$T_{1}$
from the preceding theorem verifies the unitarity condition
$$
T_{1}(x)^{\dagger} = T_{1}(x)
$$
if and only if the constants 
$f_{abc},~ f'_{abc},~f"_{abc},~h_{ab}$
and
$h'_{ab}$
have real values.
\label{uni-T1}
\end{prop}

The proof is very simple and relies on the relations (\ref{conjugate-YM}). 
To study the causality axiom in the first order of the perturbation theory, one
has to investigate some causal distributions and some relations between them. 
We have

\begin{prop}
The following distributions are well defined and have causal support:
\begin{eqnarray}
D_{m_{a}m_{b}}(x) \equiv 
D^{(+)}_{m_{a}}(x) D^{(+)}_{m_{b}}(x) - (+ \rightarrow -),
\nonumber \\
D_{m_{a}m_{b};\mu\nu} \equiv \left[ 
D^{(+)}_{m_{a}}(x) {\partial^{2} \over \partial x^{\mu} \partial x^{\nu}} 
D^{(+)}_{m_{b}}(x) - {\partial \over \partial x^{\mu} } D^{(+)}_{m_{a}}(x) 
{\partial \over \partial x^{\nu}}D^{(+)}_{m_{b}}(x) + (a \leftrightarrow b)
\right] - (+ \rightarrow -),
\nonumber \\
D_{m_{a}m_{b};\mu} \equiv 
{\partial \over \partial x^{\mu}} D^{(+)}_{m_{a}}(x) D^{(+)}_{m_{b}}(x) 
+ (+ \rightarrow -),
\nonumber \\
D_{m_{a}m_{b}m_{c}}(x) \equiv 
D^{(+)}_{m_{a}}(x) D^{(+)}_{m_{b}}(x)D^{(+)}_{m_{c}}(x) + (+ \rightarrow -),
\nonumber \\
D_{m_{a}m_{b};m_{c}}(x) \equiv 
\left[ \partial_{\mu}D^{(+)}_{m_{a}}(x) 
\partial^{\mu}D^{(+)}_{m_{b}}(x) \right] D^{(+)}_{m_{c}}(x) 
+ (+ \rightarrow -),
\nonumber \\
D_{m_{a}m_{b}m_{c}m_{d}}(x) \equiv 
D^{(+)}_{m_{a}}(x) D^{(+)}_{m_{b}}(x) D^{(+)}_{m_{c}}(x) D^{(+)}_{m_{d}}(x) 
- (+ \rightarrow -). \quad
\label{distributions}
\end{eqnarray}

Moreover, they verify the following relations:
\begin{eqnarray}
{\partial \over \partial x_{\nu}} D_{m_{a}m_{b};\mu\nu} = 
(m_{b}^{2} - m_{a}^{2}) \left[ D_{m_{a}m_{b};\mu} - (a \leftrightarrow b) 
\right], \quad
\nonumber \\
{\partial \over \partial x_{\mu}} D_{m_{a}m_{b};\mu} = 
{1\over 2} (\square + m_{b}^{2} - m_{a}^{2}) D_{m_{a}m_{b}}.
\label{D-mu-nu}
\end{eqnarray}
\end{prop}

Finally we have
\begin{prop}
The expression 
$T_{1}$
determined in the preceding theorem verifies the causality condition:
$$
[T_{1}(x), T_{1}(y)] = 0,\quad \forall x, y \in \R^{4} \quad {\rm s.t.} \quad
(x-y)^{2} < 0. 
$$
\end{prop}

One must determine the commutator appearing in the lefthand side. The
computations are similar with the one from \cite{Gr1} and we do not give them
here. We only mention that the commutator involves the distributions listed 
in (\ref{distributions}) which have causal support.

We can give now a generic form for the distribution
$T_{2}$.
We split causally the commutator 
$[T_{1}(x), T_{1}(y)]$
according to the prescription of Epstein and Glaser and include the most
general finite arbitrariness of the decomposition taking into account general
considerations explained in \cite{Gr1}. First we note that we have:
\begin{prop}
The distributions listed in (\ref{distributions}) admit causal splittings which
preserve Lorentz covariance. Moreover, the splitting can be chosen such that it
will preserve the properties (\ref{D-mu-nu}) i.e. we can arrange it such that
we have:
\begin{eqnarray}
{\partial \over \partial x_{\nu}} D^{ret(adv)}_{m_{a}m_{b};\mu\nu} = 
(m_{b}^{2} - m_{a}^{2}) \left[ D^{ret(adv)}_{m_{a}m_{b};\mu} - 
(a \leftrightarrow b) 
\right],
\nonumber \\
{\partial \over \partial x_{\mu}} D^{ret(adv)}_{m_{a}m_{b};\mu} = 
{1\over 2} (\square + m_{b}^{2} - m_{a}^{2}) D^{ret(adv)}_{m_{a}m_{b}}.
\end{eqnarray}
\label{canonical-split}
\end{prop}

So we can provide now the generic expression of the distribution
$T_{2}$.. 
The expression is extremely long, but we provide it because it provides the
easiest way to compute explicit effects in a concrete theory, like the standard
model. (For this, one should include, of course, the lepton fields). We will 
denote by
$
D^{F}_{m}(x),~D^{F}_{m_{a}m_{b}}(x), ~D^{F}_{m_{a}m_{b};\mu\nu},~
D^{F}_{m_{a}m_{b};\mu},~D^{F}_{m_{a}m_{b}m_{c}}(x),~D^{F}_{m_{a}m_{b};m_{c}}(x)
$
etc. the corresponding Feynman propagators and observe that they verify
equations of the same type as those from the preceding proposition. We have by
long and tedious computations:
\begin{prop}
The generic form of the distribution 
$T_{2}$
is 
\be
T_{2}(x,y) = :T_{1}(x) T_{1}(y): + T^{0}_{2}(x,y) + T^{h}_{2}(x,y) 
+ \delta(x-y) L(x)
\ee
where
\begin{eqnarray}
T^{0}_{2}(x,y) = 
- f_{cab} f_{cde} D_{m_{c}}^{F}(x-y) \{ 
:A_{a\nu}(x) F_{b}^{\nu\mu}(x) A_{d}^{\rho}(y) F_{e\rho\mu}(y): 
\nonumber \\
+ :u_{a}(x) \partial_{\mu}\tilde{u}_{b}(x) 
u_{d}(y) \partial^{\mu}\tilde{u}_{e}(y): - \left[ 
A_{a\nu}(x) F_{b}^{\nu\mu}(x) u_{d}(y) \partial_{\mu}\tilde{u}_{e}(y): 
+ (x \leftrightarrow y) \right] \}
\nonumber \\
+ {1\over 2} f_{abc} f_{dbc} D_{m_{b}m_{c}}^{F}(x-y) 
:F_{a}^{\nu\mu}(x) F_{d\nu\mu}(y):
\nonumber \\
+ f_{cab} f_{cde} {\partial \over \partial x^{\rho}} D_{m_{c}}^{F}(x-y) 
\{[:A_{a\nu}(x) F_{b}^{\nu\mu}(x) A_{d\mu}(y) A_{e}^{\rho}(y): 
+ :A_{a}^{\mu}(x) A_{b}^{\rho}(x) u_{d}(y) \partial_{\mu}\tilde{u}_{e}(y): 
\nonumber \\
+ :A_{a}^{\mu}(x) \partial_{\mu}\tilde{u}_{b}(x) A_{d}^{\rho}(y) u_{e}(y): ]
- (x \leftrightarrow y) \}
\nonumber \\
+ f_{cab} f_{cde} {\partial^{2} \over \partial x^{\mu} \partial x^{\nu}}
D_{m_{c}}^{F}(x-y) 
:A_{a\rho}(x) A_{b}^{\mu}(x) A_{d}^{\rho}(y) A_{e}^{\nu}(y): 
\nonumber \\
+ f_{abc} f_{dbc} \left[ m_{b}^{2} g_{\mu\nu} D_{m_{b}m_{c}}^{F}(x-y) 
- 2 D_{m_{b}m_{c};\mu\nu}^{F}(x-y) \right] : A_{a}^{\mu}(x) A_{d}^{\nu}(y):
\nonumber \\
+ f_{abc} f_{dbc} D_{m_{b},m_{c};\rho}^{F}(x-y) 
\{ \left[ :A_{a\mu}(x) F_{d}^{\mu\rho}(y): 
+ :u_{a}(x) \partial^{\rho}\tilde{u}_{d}(y): \right] - (x \leftrightarrow y)\}
\nonumber \\
-  f_{abc} f_{abc} \left({1\over 3}\square - 2m_{a}^{2}\right) 
D_{m_{a}m_{b}m_{c}}^{F}(x-y) {\bf 1} 
\nonumber \\
+ f_{cab} f'_{dec} D_{m_{c}}^{F}(x-y) \{ [
:A_{a\mu}(x) F_{b}^{\mu\nu}(x) \Phi_{d}(y) \partial_{\nu}\Phi_{e}(y): 
\nonumber \\
+ :u_{a}(x) \partial_{\mu}\tilde{u}_{b}(x) \Phi_{d}(y) 
\partial^{\mu}\Phi_{e}(y):] + (x \leftrightarrow y)\}
\nonumber \\
+ f_{cab} f'_{dec} {\partial \over \partial x^{\rho}} D_{m_{c}}^{F}(x-y) 
\left[ :A_{a}^{\mu}(x) A_{b}^{\rho}(x) \Phi_{d}(y) \partial_{\mu}\Phi_{e}(y): 
- (x \leftrightarrow y) \right]
\nonumber \\
- 2 f_{abc} h^{(1)}_{dec} D_{m_{c}}^{F}(x-y) \{
[:A_{a\mu}(x) F_{b}^{\mu\nu}(x) \Phi_{d}(y) A_{e\nu}(y): 
\nonumber \\
- :u_{a}(x) \partial_{\mu}\tilde{u}_{b}(x) \Phi_{d}(y) A_{e}^{\mu}(y):] 
+ (x \leftrightarrow y)\}
\nonumber \\
-2 f_{abc} h^{(1)}_{dec} {\partial \over \partial x^{\rho}} D_{m_{c}}^{F}(x-y) 
\left[ :A_{a\mu}(x) A_{b}^{\rho}(x) \Phi_{d}(y) A_{e}^{\mu}(y): 
- (x \leftrightarrow y) \right]
\nonumber \\
- f_{cab} h^{(2)}_{dce} D_{m_{c}}^{F}(x-y) \left[
:A_{a}^{\mu}(x) \partial_{\mu}\tilde{u}_{b}(x) \Phi_{d}(y) u_{e}(y): 
 + (x \leftrightarrow y)\right]
\nonumber \\
- f_{cab} h^{(2)}_{dec} {\partial \over \partial x^{\rho}} D_{m_{c}}^{F}(x-y) 
\left[ :A_{a}^{\rho}(x) u_{b}(x) \Phi_{d}(y) \tilde{u}_{e}(y): -
(x \leftrightarrow y) \right]
\nonumber \\
- f_{cab} h^{(2)}_{dcb} D_{m_{b}m_{c};\mu}^{F}(x-y) \left[
:A_{a}^{\mu}(x) \Phi_{d}(y): - (x \leftrightarrow y)\right]
\nonumber \\
+ f'_{cab} f'_{cde} D_{m^{*}_{c}}^{F}(x-y) 
:\partial_{\mu}\Phi_{a}(x) A_{b}^{\mu}(x) 
\partial_{\nu}\Phi_{d}(y) A_{e}^{\nu}(y): 
\nonumber \\
+ f'_{cab} f'_{cde} [{\partial \over \partial x^{\rho}} D_{m^{*}_{c}}^{F}(x-y) 
:\partial_{\mu}\Phi_{a}(x) A_{b}^{\mu}(x) \Phi_{d}(y) A_{e}^{\rho}(y): 
\nonumber \\
- {\partial \over \partial x^{\mu}} D_{m^{*}_{c}}^{F}(x-y) 
:\Phi_{a}(x) A_{b}^{\mu}(x) \partial_{\rho}\Phi_{d}(y) A_{e}^{\rho}(y):]
\nonumber \\
- f'_{cab} f'_{cde} {\partial \over \partial x^{\mu}\partial x^{\nu}} 
D_{m^{*}_{c}}^{F}(x-y) 
:\Phi_{a}(x) A_{b}^{\mu}(x) \Phi_{d}(y) A_{e}^{\nu}(y): 
\nonumber \\
- f'_{abc} f'_{dec} D_{m_{c}}^{F}(x-y) 
:\Phi_{a}(x) \partial_{\mu}\Phi_{b}(x) 
\Phi_{d}(y) \partial^{\mu}\Phi_{e}(y): 
\nonumber \\
- f'_{abc} f'_{abd} D_{m^{*}_{a}m^{*}_{b};\mu\nu}^{F}(x-y) 
:A_{c}^{\mu}(x) A_{d}^{\nu}(x): 
\nonumber \\
- f'_{abc} f'_{adc} D_{m^{*}_{a}m_{c}}^{F}(x-y) 
:\partial_{\mu}\Phi_{b}(x) \partial^{\mu}\Phi_{d}(x): 
\nonumber \\
- f'_{abc} f'_{adc} D_{m_{c}m^{*}_{a};\mu}^{F}(x-y) 
\left[ :\partial^{\mu}\Phi_{b}(x) \Phi_{d}(x): - (x \leftrightarrow y) \right]
\nonumber \\
- f'_{abc} f'_{adc} (m^{*}_{a})^{2} D_{m^{*}_{a}m_{c}}^{F}(x-y) 
:\Phi_{b}(x) \Phi_{d}(x): 
\nonumber \\
- f'_{abc} f'_{abc} \left[ D_{m^{*}_{a}m^{*}_{b};m_{c}}^{F}(x-y) 
+ (m^{*}_{a})^{2} D_{m^{*}_{a}m^{*}_{b}m_{c}}^{F}(x-y)\right] {\bf 1}
\nonumber \\
- 2 f'_{abc} h^{(1)}_{dec} D_{m_{c}}^{F}(x-y) 
\left[:\Phi_{a}(x) \partial_{\mu}\Phi_{b}(x) \Phi_{d}(y) A_{e}^{\mu}(y): 
+ (x \leftrightarrow y)\right]
\nonumber \\
+ f'_{cab} h^{(1)}_{cde} D_{m^{*}_{c}}^{F}(x-y) 
\left[:\partial_{\mu}\Phi_{a}(x) A_{b}^{\mu}(x) A_{d\rho}(y) A_{e}^{\rho}(y): 
+ (x \leftrightarrow y)\right]
\nonumber \\
- f'_{cab} h^{(1)}_{cde} {\partial \over \partial x^{\mu}} 
D_{m^{*}_{c}}^{F}(x-y) 
\left[:\Phi_{a}(x) A_{b}^{\mu}(x) A_{d\rho}(y) A_{e}^{\rho}(y): 
- (x \leftrightarrow y)\right]
\nonumber \\
+ 2 f'_{abc} h^{(1)}_{bcd} D_{m^{*}_{b}m_{c}}^{F}(x-y) 
\left[:\partial_{\mu}\Phi_{a}(x) A_{d}^{\mu}(y): 
+ (x \leftrightarrow y)\right]
\nonumber \\
- 2 f'_{abc} h^{(1)}_{bcd} D_{m_{c}m^{*}_{b};\mu}^{F}(x-y) 
\left[:\Phi_{a}(x) A_{d}^{\mu}(y): 
- (x \leftrightarrow y)\right]
\nonumber \\
+ f'_{cab} h^{(2)}_{cde} D_{m^{*}_{c}}^{F}(x-y) 
\left[ :\partial_{\mu}\Phi_{a}(x) A_{b}^{\mu}(x) \tilde{u}_{d}(y) u_{e}(y): 
+ (x \leftrightarrow y)\right]
\nonumber \\
- f'_{cab} h^{(2)}_{cde} {\partial \over \partial x^{\rho}} 
D_{m^{*}_{c}}^{F}(x-y) 
\left[ :\Phi_{a}(x) A_{b}^{\rho}(x) \tilde{u}_{d}(y) u_{e}(y): 
- (x \leftrightarrow y)\right]
\nonumber \\
+ 3f'_{cab} f^{"}_{cde} D_{m^{*}_{c}}^{F}(x-y) 
\left[ :\partial_{\mu}\Phi_{a}(x) A_{b}^{\mu}(x) \Phi_{d}(y) \Phi_{e}(y): 
+ (x \leftrightarrow y)\right]
\nonumber \\
- 3f'_{cab} f^{"}_{cde} 
{\partial \over \partial x^{\rho}} D_{m^{*}_{c}}^{F}(x-y) 
\left[ :\Phi_{a}(x) A_{b}^{\rho}(x) \Phi_{d}(y) \Phi_{e}(y): 
- (x \leftrightarrow y)\right]
\nonumber \\
+ 6f'_{bca} f^{"}_{bcd} D_{m^{*}_{c}m^{*}_{b};\rho}^{F}(x-y) 
\left[ :A_{x}^{\rho}(x) \Phi_{d}(y): - (x \leftrightarrow y)\right]
\nonumber \\
+ 4f'_{cab} g_{cdef} D_{m^{*}_{c}}^{F}(x-y) 
\left[ :\partial_{\mu}
\Phi_{a}(x) A_{b}^{\mu}(x) \Phi_{d}(y) \Phi_{e}(y) \Phi_{f}(y): 
+ (x \leftrightarrow y)\right]
\nonumber \\
- 4f'_{cab} g_{cdef} {\partial \over \partial x^{\rho}} D_{m^{*}_{c}}^{F}(x-y) 
\left[ :\Phi_{a}(x) A_{b}^{\rho}(x) \Phi_{d}(y) \Phi_{e}(y) \Phi_{f}(y): 
- (x \leftrightarrow y)\right]
\nonumber \\
- 12f'_{bca} g_{bcde} D_{m^{*}_{b}m^{*}_{c};\rho}^{F}(x-y) 
\left[ :A_{a}^{\rho}(x) \Phi_{d}(y) \Phi_{e}(y): 
- (x \leftrightarrow y)\right]
\nonumber \\
+ h^{(1)}_{cab} h^{(1)}_{cde} D_{m^{*}_{c}}^{F}(x-y) 
:A_{a\mu}(x) A_{b}^{\mu}(x) A_{d\nu}(y) A_{e}^{\nu}(y): 
\nonumber \\
- 4h^{(1)}_{cab} h^{(1)}_{dbe} D_{m_{b}}^{F}(x-y) 
:\Phi_{a}(x) A_{c}^{\mu}(x) \Phi_{d}(y) A_{e\mu}(y): 
\nonumber \\
- 4h^{(1)}_{abc} h^{(1)}_{abd} D_{m^{*}_{a}m_{b}}^{F}(x-y) 
:A_{c}^{\mu}(x) A_{d\mu}(y): 
+ 2h^{(1)}_{abc} h^{(1)}_{dbc} D_{m_{b}m_{c}}^{F}(x-y) 
:\Phi_{a}(x) \Phi_{d}(y): 
\nonumber \\
+ 2h^{(1)}_{abc} h^{(1)}_{abc} D_{m^{*}_{a}m_{b}m_{c}}^{F}(x-y) {\bf 1}
\nonumber \\
+ h^{(1)}_{cab} h^{(2)}_{cde} D_{m^{*}_{c}}^{F}(x-y) 
\left[ :A_{a\mu}(x) A_{b}^{\mu}(x) \tilde{u}_{d}(y) u_{e}(y): 
+ (x \leftrightarrow y)\right]
\nonumber \\
+ 3h^{(1)}_{cab} h^{(3)}_{cde} D_{m^{*}_{c}}^{F}(x-y) 
\left[ :A_{a\mu}(x) A_{b}^{\mu}(x) \Phi_{d}(y) \Phi_{e}(y): 
+ (x \leftrightarrow y)\right]
\nonumber \\
+ 4h^{(1)}_{cab} g_{cdef} D_{m^{*}_{c}}^{F}(x-y) 
\left[ :A_{a\mu}(x) A_{b}^{\mu}(x) \Phi_{d}(y) \Phi_{e}(y) \Phi_{f}(y): 
+ (x \leftrightarrow y)\right]
\nonumber \\
+ h^{(2)}_{abc} h^{(2)}_{ade} D_{m^{*}_{a}}^{F}(x-y) 
:\tilde{u}_{b}(x) u_{c}(x) \tilde{u}_{d}(y) u_{e}(y): 
\nonumber \\
+ h^{(2)}_{abc} h^{(2)}_{deb} D_{m_{b}}^{F}(x-y) 
\left[ :\Phi_{a}(x) u_{c}(x) \Phi_{d}(y) \tilde{u}_{e}(y): 
- (x \leftrightarrow y)\right]
\nonumber \\
+ h^{(2)}_{abc} h^{(2)}_{aeb} D_{m^{*}_{a}m_{b}}^{F}(x-y) 
\left[ :\tilde{u}_{e}(x) u_{c}(y): + (x \leftrightarrow y)\right]
\nonumber \\
- h^{(2)}_{abc} h^{(2)}_{dbc} D_{m_{b}m_{c}}^{F}(x-y) 
:\Phi_{a}(x) \Phi_{d}(y): 
- h^{(2)}_{abc} h^{(2)}_{acb} D_{m^{*}_{a}m_{b}m_{c}}^{F}(x-y) {\bf 1}
\nonumber \\
+ 3h^{(2)}_{cab} f^{"}_{cde} D_{m^{*}_{c}}^{F}(x-y) 
\left[ :\tilde{u}_{a}(x) u_{b}(x) \Phi_{d}(y) \Phi_{e}(y): 
+ (x \leftrightarrow y)\right]
\nonumber \\
+ 4h^{(2)}_{cab} g_{cdef} D_{m^{*}_{c}}^{F}(x-y) 
\left[ :\tilde{u}_{a}(x) u_{b}(x) \Phi_{d}(y) \Phi_{e}(y) \Phi_{f}(y): 
+ (x \leftrightarrow y)\right]
\nonumber \\
+ 9f{"}_{cab} f^{"}_{cde} D_{m^{*}_{c}}^{F}(x-y) 
:\Phi_{a}(x) \Phi_{b}(x) \Phi_{d}(y) \Phi_{e}(y): 
\nonumber \\
+ 18f^{"}_{abc} f^{"}_{abd} D_{m^{*}_{a}m^{*}_{b}}^{F}(x-y) 
:\Phi_{c}(x) \Phi_{d}(y):
+ 18f^{"}_{abc} f^{"}_{abc} D_{m^{*}_{a}m^{*}_{b}m^{*}_{c}}^{F}(x-y) {\bf 1}
\nonumber \\
+ 12f{"}_{cab} g_{cdef} D_{m^{*}_{c}}^{F}(x-y) 
\left[ :\Phi_{a}(x) \Phi_{b}(x) \Phi_{d}(y) \Phi_{e}(y) \Phi_{f}(y): 
+ (x \leftrightarrow y)\right]
\nonumber \\
+ 18f^{"}_{abc} g_{bcde} D_{m^{*}_{b}m^{*}_{c}}^{F}(x-y) 
\left[ :\Phi_{a}(x) \Phi_{d}(y) \Phi_{e}(y):
+ (x \leftrightarrow y)\right]
\nonumber \\
+ 24f^{"}_{bcd} g_{abcd} D_{m^{*}_{b}m^{*}_{c}m^{*}_{d}}^{F}(x-y) 
\left[ \Phi_{a}(x) + + (x \leftrightarrow y)\right]
\nonumber \\
+ 16g_{abcd} g_{defh} D_{m^{*}_{d}}^{F}(x-y) 
:\Phi_{a}(x) \Phi_{b}(x) \Phi_{c}(x) \Phi_{e}(y) \Phi_{f}(y) \Phi_{h}(y): 
\nonumber \\
+ 144g_{abcd} g_{cdef} D_{m^{*}_{c}m^{*}_{d}}^{F}(x-y) 
:\Phi_{a}(x) \Phi_{b}(x) \Phi_{e}(y) \Phi_{f}(y):
\nonumber \\
+ 576g_{abcd} g_{bcde} D_{m^{*}_{b}m^{*}_{c}m^{*}_{d}}^{F}(x-y) 
:\Phi_{a}(x) \Phi_{e}(y):
+ 24g_{abcd} g_{abcd} D_{m^{*}_{a}m^{*}_{b}m^{*}_{c}m^{*}_{d}}^{F}(x-y) 
{\bf 1}. \qquad
\label{T2-generic}
\end{eqnarray}
\begin{eqnarray}
T^{h}_{2}(x,y) = 
- 2f_{cab} h_{cd} D_{m_{c}}^{F}(x-y) 
\{[:A_{a\nu}(x) F_{b}^{\nu\mu}(x) A_{d\mu}(y): 
- :u_{a}(x) \partial_{\mu}\tilde{u}_{b}(x) A_{d}^{\mu}(y): 
\nonumber \\
- :A_{a}^{\mu}(x) \partial_{\mu}\tilde{u}_{b}(y) u_{d}(y):]
+ (x \leftrightarrow y)] \}
\nonumber \\
- 2f_{cab} h_{cd} {\partial \over \partial x^{\rho}} D_{m_{c}}^{F}(x-y) 
\{[:A_{a\mu}(x) A_{b}^{\rho}(x) A_{d}^{\mu}(y):
- :A_{a}^{\rho}(x) u_{b}(x) \tilde{u}_{d}(y):]
- (x \leftrightarrow y)] \}
\nonumber \\
- 2f'_{abc} h_{cd} D_{m_{c}}^{F}(x-y) 
\left[ :\Phi_{a}(x) \partial_{\mu}\Phi_{b}(x) A_{d}^{\mu}(y): 
+ (x \leftrightarrow y)\right] 
\nonumber \\
+ 2f'_{abc} h'_{cd} D_{m^{*}_{c}}^{F}(x-y) 
\left[ :\partial_{\mu}\Phi_{a}(x) A_{b}^{\mu}(x) \Phi_{d}(y): 
+ (x \leftrightarrow y)\right] 
\nonumber \\
- 2f'_{cab} h'_{cd} 
{\partial \over \partial x^{\rho}} D_{m^{*}_{c}}^{F}(x-y) 
\left[ :\Phi_{a}(x) A_{c}^{\rho}(x) \Phi_{d}(y):
- (x \leftrightarrow y)\right] 
\nonumber \\
+ 2f'_{bca} h'_{bc} D_{m^{*}_{c}m^{*}_{b};\rho}^{F}(x-y) 
\left[ A_{a}^{\rho}(x) - (x \leftrightarrow y)\right] 
\nonumber \\
- 4h^{(1)}_{cab} h_{cd} D_{m_{c}}^{F}(x-y) 
\left[ :\Phi_{a}(x) A_{b}^{\mu}(x) A_{d\mu}(y): 
+ (x \leftrightarrow y)\right] 
\nonumber \\
+ 16 h^{(1)}_{cab} h_{cd} D_{m_{c}}^{F}(x-y) \Phi_{a}(x)
+ 2h^{(1)}_{cab} h'_{cd} D_{m^{*}_{c}}^{F}(x-y) 
\left[ :A_{a\mu}(x) A_{b}^{\mu}(x) \Phi_{d}(y): 
+ (x \leftrightarrow y)\right] 
\nonumber \\
+ 2h^{(2)}_{acb} h_{cd} D_{m_{c}}^{F}(x-y) 
\left[ :\Phi_{a}(x) u_{b}(x) \tilde{u}_{d}(y): 
+ (x \leftrightarrow y)\right] 
\nonumber \\
- 2h^{(2)}_{abc} h_{cd} D_{m_{c}}^{F}(x-y) 
\left[ :\Phi_{a}(x) \tilde{u}_{b}(x) u_{d}(y):
+ (x \leftrightarrow y)\right] 
\nonumber \\
+ 4h^{(2)}_{abc} h_{bc} D_{m_{b}m_{c}}^{F}(x-y) \Phi_{a}(x)
+ 2h^{(2)}_{cab} h'_{cd} D_{m^{*}_{c}}^{F}(x-y) 
\left[ :\tilde{u}_{a}(x) u_{b}(x) \Phi_{d}(y): 
+ (x \leftrightarrow y)\right] 
\nonumber \\
+ 6h^{(3)}_{cab} h'_{cd} D_{m^{*}_{c}}^{F}(x-y) 
\left[ :\Phi_{a}(x) \Phi_{b}(x) \Phi_{d}(y): 
+ (x \leftrightarrow y)\right] 
+ 6h^{(3)}_{abc} h'_{bc} D_{m^{*}_{b}m^{*}_{c}}^{F}(x-y) 
\Phi_{a}(x)
\nonumber \\
+ 8g_{abcd} h'_{de} D_{m^{*}_{d}}^{F}(x-y) 
\left[ :\Phi_{a}(x) \Phi_{b}(x) \Phi_{c}(x) \Phi_{e}(y): 
+ (x \leftrightarrow y)\right] 
\nonumber \\
+ 24g_{abcd} h'_{cd} D_{m^{*}_{c}m^{*}_{d}}^{F}(x-y) 
\left[ :\Phi_{a}(x)\Phi_{b}(x):
+ (x \leftrightarrow y)\right] 
\nonumber \\
-4h_{ab} h_{ab} D_{m_{c}}^{F}(x-y) \{
:A_{b}^{\mu}(x) A_{b\mu}(y): - [:\tilde{u}_{a}(x) u_{b}(y): 
+ (x \leftrightarrow y) ]\}
\nonumber \\
+ 4h_{ab} h_{ab} D_{m_{a}m_{b}}^{F}(x-y) {\bf 1} 
+ 4h'_{ac} h'_{bc} D_{m^{*}_{c}}^{F}(x-y) 
:\Phi_{a}(x) \Phi_{b}(y): 
\nonumber \\
+ 4h'_{ab} h'_{ab} D_{m^{*}_{a}m^{*}_{b}}^{F}(x-y) {\bf 1}  \qquad
\label{T2-h}
\end{eqnarray}
and
$L(x)$
is a finite renormalisation Lorentz invariant and of power 
$\leq 4$ 
i.e. a sum of terms of the type (\ref{list2}) - (\ref{list4}).
\end{prop}
\newpage
\subsection{Second Order Gauge Invariance}

We are not guaranteed that the generic expression of
$T_{2}(x,y)$
from the preceding proposition leads to a well-defined operator on the factor
space
${\cal H}^{r}_{YM}$;
as in \cite{Gr1}, one can show that this can happen if and only if some severe
restrictions are placed on the constants appearing in the expression of the
interaction Lagrangian. In \cite{ASD3} it is proved that, in the standard
model, one can choose conveniently the finite normalisation
$L(x)$
such that gauge invariance is valid in the second order of perturbation theory
(this in turn guarantees that the factorisation of the $S$-matrix is possible
in this order). We detail below this result in a more general context, when the
characteristics of the standard model are not used into the computations i.e.
we do not take specific expressions for the constants
$f_{abc}$.
As in \cite{Gr1} we observe that the generic expression for the second-order
$S$-matrix obtained in the preceding proposition corresponds to a ``canonical"
causal splitting of the commutator
$D_{2}(x,y)$;
namely, one splits causally the numerical distributions in the expression of
the commutator by making the replacements:
$D_{m} \rightarrow D_{m}^{ret(adv)}, 
\partial _{\mu} D_{m} \rightarrow \partial_{\mu} D_{m}^{ret(adv)},
D_{m_{a}m_{b};\mu\nu} \rightarrow D^{ret(adv)}_{m_{a}m_{b};\mu\nu},
$
etc. In this way one obtains the expressions
$R^{0}_{2}(x,y)$
and
$A^{0}_{2}(x,y)$
which will be called the {\it canonical causal splitting}. This splitting leads 
to the expression
$T^{0}_{2}(x,y) + T^{h}_{2}(x,y)$
from the preceding proposition. Now we have:

\begin{thm}
The expression 
$T_{2}$
appearing in the preceding proposition leads, in the adiabatic limit, to a well
defined operator on 
${\cal H}^{r}_{YM}$ 
if and only if the following identities are verified:

\be
f_{abc}f_{dec} + f_{bdc} f_{aec} + f_{dac} f_{bec} = 0, 
\quad a, b, d, e = 1,\dots,r;
\label{Jacobi}
\ee
\be
f'_{dca} f'_{ceb} - f'_{dcb} f'_{cea} = - f_{abc} f'_{dec},
\quad a, b, d, e = 1,\dots,r;
\label{repr-f'}
\ee
\be
f'_{cab} f^{"}_{cde} + f'_{cdb} f^{"}_{cae} + f'_{ceb} f^{"}_{cda} = 0,
\quad {\rm iff} \quad m_{a} = m_{b} = m_{d} = m_{e} = 0;
\label{h3-0}
\ee
\be
f_{abc} h_{cd} = 0;
\label{f-h}
\ee
\be
f'_{dba} h'_{cd} + f'_{dca} h'_{bd} = 0;
\label{h3-2}
\ee
\be
f'_{cba} g_{cdef} +
(b \leftrightarrow d) + (b \leftrightarrow e) + (b \leftrightarrow f) = 0,
\quad {\rm iff} \quad m_{a} = m_{b} = m_{d} = m_{e} = m_{f} = 0.
\label{f'g}
\ee
\label{T2}
\end{thm}

{\bf Proof:}
(i) We follow the ideas from \cite{DHKS1} and \cite{Gr1}. We first have 
\be
d_{Q} D_{2}(x,y) = 
i {\partial \over \partial x^{\mu}} [T^{\mu}_{1}(x), T_{1}(y)] -
(x \leftrightarrow y)
\label{split-d2-a}
\ee
and we must compute the right hand side. It is elementary to see that the
distribution
$d_{Q} D_{2}(x,y)$
still has a causal support so it can be split causally:
\be
d_{Q} D_{2}(x,y) = d_{Q} A_{2}(x,y) - d_{Q} R_{2}(x,y).
\label{split-d2-b}
\ee

If we split causally the right hand side of the formula (\ref{split-d2-a})
preserving Lorentz covariance and power counting, we will obviously get 
valid expressions for the distributions
$d_{Q} A_{2}(x,y)$
and
$d_{Q} R_{2}(x,y)$.
If we want to obtain exactly 
$d_{Q} A^{0}_{2}(x,y)$
and
$d_{Q} R^{0}_{2}(x,y)$
we must compute the commutators in the (\ref{split-d2-a}), next perform the
derivatives and finally extract the canonical causal splitting.  Of course, in
this way we do not get the most general expression for these distribution
because we have the possibility of finite normalisations. But the arbitrariness
for
$d_{Q} R_{2}(x,y)$
is {\it exactly the same} as the arbitrariness for
$d_{Q} T_{2}(x,y)$
i.e. of the form
$\delta(x-y) N(x)$.
So, we get the most general expression for the distributions
$d_{Q} A_{2}(x,y)$
and
$d_{Q} R_{2}(x,y)$.

Because we have (\ref{gi-epsilon}) for 
$n = 1$, 
we conclude that (\ref{gi-epsilon}) for $n = 2$ is equivalent to:
\be
\int_{\R^{4} \times \R^{4}} dx dy g_{\epsilon}(x) g_{\epsilon}(y)
\left. d_{Q} R_{2}(x,y)\right|_{Ker(Q)} = O(\epsilon).
\label{gi-t2}
\ee

Imposing this condition on the expression determined in the way outlined above
will lead to the conditions from the statement.

(ii) By straightforward computation we obtain the following expression for the
first commutator appearing in (\ref{split-d2-a}). We obtain
\begin{eqnarray}
[T^{\mu}_{1}(x), T_{1}(y)] = 
{\partial \over \partial x^{\mu}} D_{m_{c}}(x-y) T_{c}(x,y) 
+ {\partial^{2} \over \partial x_{\mu}\partial x^{\rho}} D_{m_{c}}(x-y) 
T_{c}^{\rho}(x,y) 
\nonumber \\
+ {\partial \over \partial x_{\mu}} D_{m^{*}_{c}}(x-y) T^{(*)}_{c}(x,y) 
+ {\partial^{2} \over \partial x_{\mu}\partial x^{\rho}} D_{m^{*}_{c}}(x-y) 
T_{c}^{(*)\rho}(x,y) + \cdots
\label{[k,t]}
\end{eqnarray}
where by $\cdots$ we mean contributions which will not produce anomalies.

We give the explicit expression of the operator-valued distributions
$T_{c}$, $T^{*}_{c}$,
etc.
\begin{eqnarray}
T_{c}(x,y) \equiv
f_{cab} f_{cde} 
[:u_{a}(x) A_{b}^{\nu}(x) A_{d}^{\rho}(y) F_{e\rho\nu}(y): 
- :u_{a}(x) A_{b}^{\nu}(x) u_{d}(y) \partial_{\nu}\tilde{u}_{e}(y): 
\nonumber \\
+ {1\over 2} 
:u_{a}(x) u_{b}(x) A_{d}^{\rho}(y) \partial_{\rho}\tilde{u}_{e}(y): ]
+ f_{abc} f'_{dec} 
:u_{a}(x) A_{b}^{\rho}(x) \Phi_{d}(y) \partial_{\rho}\Phi_{e}(y): 
\nonumber \\
+ 2f_{abc} h^{(1)}_{dec} 
:u_{a}(x) A_{b\rho}(x) \Phi_{d}(y) A_{e}^{\rho}(y): 
- {1\over 2} f_{abc} h^{(2)}_{dec} 
:u_{a}(x) u_{b}(x) \Phi_{d}(y) \tilde{u}_{e}(y): 
\nonumber \\
+ f_{cab} h_{cd} [2 :u_{a}(x) A_{b\nu}(x) A_{d}^{\nu}(y):
+ :u_{a}(x) u_{b}(x) \tilde{u}_{d}(y):]  
\end{eqnarray}
\begin{eqnarray}
T^{(*)}_{c}(x,y) \equiv
- f'_{cab} f'_{cde} 
:\Phi_{a}(x) u_{b}(x) \partial_{\rho}\Phi_{d}(y) A_{e}^{\rho}(y): 
- f'_{cab} h^{(1)}_{cde} 
:\Phi_{a}(x) u_{b}(x) A_{d\rho}(y) A_{e}^{\rho}(y): 
\nonumber \\
- f'_{cab} h^{(2)}_{cde} 
:\Phi_{a}(x) u_{b}(x) \tilde{u}_{d}(y) u_{e}(y): 
- 3f'_{cab} f^{"}_{cde} 
:\Phi_{a}(x) u_{b}(x) \Phi_{d}(y) \Phi_{e}(y): 
\nonumber \\
- 4f'_{acb} g_{cdef} 
:\Phi_{a}(x) u_{b}(x) \Phi_{d}(y) \Phi_{e}(y) \Phi_{f}(y): 
- 2f'_{cab} h'_{cd} :\Phi_{a}(x) u_{b}(x) \Phi_{d}(y): \quad
\end{eqnarray}
and
\be
T_{c}^{\rho}(x,y) \equiv
- f_{cab} f_{cde} 
:u_{a}(x) A_{b\nu}(x) A_{d}^{\nu}(y) A_{e}^{\rho}(y): 
\ee
\be
T_{c}^{(*)\rho}(x,y) \equiv
- f'_{cab} f'_{cde} 
:\Phi_{a}(x) u_{b}(x) \Phi_{d}(y) A_{e}^{\rho}(y): .
\ee

(iii) We must split causally the distribution
$
{\partial \over \partial x^{\mu}} [T^{\mu}_{1}(x), T_{1}(y)].
$
It is important that the
causal splitting can be done in such a way that we have the relations from
proposition \ref{canonical-split}. This is the reason that terms of the type
$D_{m_{b}m_{c};\rho}(x-y) T^{\mu\rho}_{bc}(x,y)$
and
$D_{m_{b}m_{c}}^{\mu\rho}(x-y) T_{bc;\rho}(x,y)$
do not produce anomalies. For other terms one can reason as in  \cite{Gr1} and
one obtains the following expression for the canonical splitting:
\be
\left( {\partial \over \partial x^{\mu}} 
[T^{\mu}_{1}(x), T_{1}(y)]\right)^{ret} =
{\partial \over \partial x^{\mu}} \tilde{R}_{1}^{\mu}(x,y) -
i \delta (x-y) A(x,y)
\ee
where 
\be
\tilde{R}_{1}^{\mu}(x,y) \equiv R_{1}^{\mu}(x,y) - i \delta (x-y)
\sum_{c=1}^{r} \left[ T_{c}^{\mu}(x,y) + T_{c}^{(*)\mu}(x,y)\right]
\ee
(with
$R_{1}^{\mu}(x,y)$
given by (\ref{[k,t]}) with 
$D_{m} \rightarrow D_{m}^{ret}$, etc.)
and the {\it anomaly} is given by:
\be
A(x,y) \equiv \sum_{c=1}^{r}\left[ 
T_{c}(x,y) - {\partial \over \partial x^{\rho}} T_{c}^{\rho}(x,y) 
+ T_{c}^{(*)}(x,y) - {\partial \over \partial x^{\rho}} T_{c}^{(*)\rho}(x,y)
\right] 
\ee

The factorisation condition (\ref{gi-t2}) can be written now as follows:
\be
\int_{\R^{4}} dx (g_{\epsilon}(x))^{2}
\left[ 2A(x)- d_{Q} L(x)\right]_{Ker(Q)} = O(\epsilon).
\ee
where 
$L(x)$
is a finite normalisation and 
$A(x) \equiv A(x,x)$.

After performing the computations and some rearrangements, the last condition
writes as follows:
\begin{eqnarray}
\int_{\R^{4}} dx g_{\epsilon}(x)^{2}
[(2 f_{cae} f_{cdb} - f_{cab} f_{cde})
:u_{a}(x) F_{b\rho\nu}(x) A_{d}^{\nu}(x) A_{e}^{\rho}(x): 
\nonumber \\
- f_{cab} f_{cbe} 
:\partial_{\rho}u_{a}(x) A_{b\nu}(x) A_{d}^{\rho}(x) A_{e}^{\nu}(x): 
\nonumber \\
+ (2 f_{cad} f_{cbe} - f_{cab} f_{cde})
:u_{a}(x) u_{b}(x) A_{d}^{\rho}(x) \partial_{\rho}\tilde{u}_{e}(x): 
\nonumber \\
+ 2(f_{abc} f'_{dec} - 2f'_{cda} f'_{ceb})
:u_{a}(x) A_{b}^{\rho}(x) \Phi_{d}(y) \partial_{\rho}\Phi_{e}(y): 
\nonumber \\
+ 2( 2f_{abc} h^{(1)}_{dec} - f'_{cda} h^{(1)}_{ceb})
:u_{a}(x) A_{b\rho}(x) \Phi_{d}(y) A_{e}^{\rho}(y): 
\nonumber \\
- (f_{abc} h^{(2)}_{dec} - 2f'_{cda} h^{(2)}_{ceb})
:u_{a}(x) u_{b}(x) \Phi_{d}(y) \tilde{u}_{e}(y): 
\nonumber \\
- 2f'_{cab} f'_{cde} 
:\Phi_{a}(x) u_{b}(x) \Phi_{d}(y) \partial_{\rho}A_{e}^{\rho}(y): 
- 6f'_{cab} f^{"}_{cde} 
:\Phi_{a}(x) u_{b}(x) \Phi_{d}(y) \Phi_{e}(y): 
\nonumber \\ 
+ 4f'_{cba} g_{cdef} 
:u_{a}(x) \Phi_{b}(x)\Phi_{d}(x) \Phi_{e}(x) \Phi_{f}(x): 
\nonumber \\
- 4 f_{cab} h_{cd} :u_{a}(x) A_{b\nu}(x) A_{d}^{\nu}(x):
- 2 f_{cab} h_{cd} :u_{a}(x) u_{b}(x) \tilde{u}_{d}(x):
\nonumber \\
\left. - 4 f'_{cab} h'_{cd} :\Phi_{a}(x) u_{b}(x) \Phi_{d}(x):
- d_{Q} L(x)]\right|_{Ker(Q)} = O(\epsilon) \quad
\label{gi-2}
\end{eqnarray}

One has to compute the expression 
$d_{Q} L(x)$
taking into account the generic form for
$L(x)$
described in the preceding Subsection. One takes into account (\ref{list2}) -
(\ref{list4}) and the corresponding expressions from \cite{Gr1}; to avoid
confusions we will append a tilde sign to all coefficients in these
expressions. Because of the presence of the term 
$
:u_{a} \Phi_{b} \Phi_{d} \Phi_{e} \Phi_{f}: 
$
in the anomaly, we will include in 
$L(x)$
an expression of the type
$
\tilde{g}^{(8)}_{abdef} :\Phi_{a} \Phi_{b} \Phi_{d} \Phi_{e} \Phi_{f}: 
$
where the expression
$\tilde{g}^{(8)}_{abdef}$
is completely symmetric in all indices. Because only the expression 
$d_{Q}L(x)$
is relevant, we can take the expressions
$\tilde{g}^{(7)}_{abde}$
and
$\tilde{g}^{(8)}_{abdef}$
to be zero when all indices correspond to zero masses.

We equate with zero the coefficients of the linearly independent
(integrated) Wick monomials; we have the following cases:
\begin{itemize}
\item
We consider the coefficients of the (linearly independent) integrated Wick 
monomials:
$
\int_{\R^{4}} dx g^{2}_{\epsilon}
:u_{a} u_{b} A_{d}^{\rho} \partial_{\rho}\tilde{u}_{e}: \quad
\int_{\R^{4}} dx g^{2}_{\epsilon}
:\partial_{\rho} u_{a} u_{b} A_{d}^{\rho} \tilde{u}_{e}: \quad
\int_{\R^{4}} dx g^{2}_{\epsilon}
:A_{a\mu} A_{b}^{\mu} u_{c} \partial_{\rho} A_{d}^{\rho}: \quad
$
and
$
\int_{\R^{4}} dx g^{2}_{\epsilon} :u_{a} A_{b}^{\mu} A_{d\nu} A_{e}^{\nu}:
$
and we get, like in \cite{Gr1}, that the constants
$
f_{abc}
$
verify Jacobi identity (\ref{Jacobi}). Moreover we obtain the expression of
$\tilde{g}^{(1)}_{abde}$:
\be
\tilde{g}^{(1)}_{abde} = {i \over 8} \left( f_{cad} f_{cbe} + f_{cae} f_{cbd}
\right). 
\label{g1}
\ee
\item
From the coefficients of the Wick monomials
$
\int_{\R^{4}} dx g^{2}_{\epsilon} :u_{a} u_{b} \tilde{u}_{d}: \quad 
\int_{\R^{4}} dx g^{2}_{\epsilon} :\Phi_{a} \partial_{\mu}A_{b}^{\mu} u_{d}:
\quad
\int_{\R^{4}} dx g^{2}_{\epsilon} :\Phi_{a} \partial^{\mu}\Phi_{b} 
\partial_{\mu}u_{d}: \quad
\int_{\R^{4}} dx g^{2}_{\epsilon} :\partial_{\nu}A_{a\mu} 
\partial^{\mu}A_{b}^{\nu} u_{d}: \quad
\int_{\R^{4}} dx g^{2}_{\epsilon} :u_{a} \partial_{\mu}A_{b\nu}
\partial^{\mu}A_{d}^{\nu}: 
$
and
$
\int_{\R^{4}} dx g^{2}_{\epsilon} :u_{a} \partial_{\mu}A_{b}^{\mu}
\partial_{\nu}A_{d}^{\nu}:
$
we obtain (\ref{f-h}).
\item
From the coefficients of the monomials
$
\int_{\R^{4}} dx g^{2}_{\epsilon} :u_{a} A_{b}^{\rho} \Phi_{d}
\partial_{\rho}\Phi_{e}: \quad 
\int_{\R^{4}} dx g^{2}_{\epsilon} :u_{a} A_{b\rho} \Phi_{d}
A_{e}^{\rho}: \quad
\int_{\R^{4}} dx g^{2}_{\epsilon} :u_{a} u_{b} \Phi_{d} \tilde{u}_{e}:\quad 
\int_{\R^{4}} dx g^{2}_{\epsilon}(x) :\Phi_{a} u_{b} \Phi_{d} \Phi_{e}: 
$
and
$
\int_{\R^{4}} dx g^{2}_{\epsilon} :\Phi_{a} \Phi_{b} 
\partial_{\mu}A_{d}^{\mu} u_{e}:
$
we obtain the following system of equations which is harder to analyse than the
previous ones:
\be
2(f_{abc} f'_{dec} - 2f'_{cda} f'_{ceb}) + 4i \tilde{g}^{(5)}_{deba} = 0,
\label{1'}
\ee
\be
\left[ ( 2f_{abc} h^{(1)}_{dec} - f'_{cda} h^{(1)}_{ceb}) +
(b \leftrightarrow e) \right] - 2i \tilde{g}^{(5)}_{adbe}m_{a} = 0,
\label{2'}
\ee
\be
\left[ (f_{abc} h^{(2)}_{dec} - 2f'_{cda} h^{(2)}_{ceb}) -
(a \leftrightarrow b) \right] + 
i \left[ \tilde{g}^{(6)}_{adeb}m_{a} - (a \leftrightarrow b) \right] = 0,
\label{3'}
\ee
\be
6f'_{cab} f^{"}_{cde} -i\tilde{g}^{(6)}_{deab} 
+ 4i \tilde{g}^{(7)}_{abde} m_{b} + {\rm cyclic~ perm~ (a,d,e)} = 0,
\label{4'}
\ee
\be
-\left( f'_{cae} f'_{cbd} + f'_{cbe} f'_{cad}\right) + 
2i \tilde{g}^{(5)}_{abde} + i \tilde{g}^{(6)}_{abde} = 0.
\label{5'}
\ee

It is very nice that one can solve explicitly this system. First, we take into
account that the constants
$\tilde{g}^{(5)}_{deba}$
are symmetric in $d$ and $e$ and also in $b$ and $a$. So, if we take the
antisymmetric part in $d$ and $e$ of the relation (\ref{1'}) we get the
relation (\ref{repr-f'}) from the statement and from the symmetric part we
obtain the explicit expression for
$\tilde{g}^{(5)}_{deba}$:
\be
\tilde{g}^{(5)}_{deba} = -{i \over 2} (f_{cda} f'_{ceb} + f'_{cdb} f'_{cea}).
\label{g5}
\ee
If we substitute this expression into the equation (\ref{5'}) we get
\be
\tilde{g}^{(6)}_{abde} = 0;
\label{g6}
\ee
then, the relation (\ref{3'}) becomes a consequence of (\ref{repr-f'}) if we
use the explicit expression (\ref{h2}) for
$h^{(2)}_{abc}$.

Next, we introduce the expressions (\ref{g5}) and (\ref{h1}) of
$\tilde{g}^{(5)}_{abde}$
and resp.
$h^{(1)}_{cab}$
into the equation (\ref{2'}) and we obtain an identity if we take into account
that the constants
$f'_{abc}$
verify the equations (\ref{f-f'}) and (\ref{repr-f'}). 

If we substitute now (\ref{g6}) into the equation (\ref{4'}) we have two cases:

(a) $m_{b} \not= 0$;

In this case we get an explicit expression for
$\tilde{g}^{(7)}$:
\be
\tilde{g}^{(7)}_{abde} = {i \over 2m_{b}} (f'_{cab} f^{"}_{cde} + 
f'_{cdb} f^{"}_{cae} +f'_{ceb} f^{"}_{cad}).
\label{g7}
\ee

Let us note that this definition is consistent: the symmetry of the right hand
side in the indices $d$ and $e$ is obvious; the symmetry in the indices $a$ 
and $b$ (for
$m_{a} \not= 0, \quad m_{b} \not= 0$)
follows from the expression (\ref{f"}) for the symbols 
$f^{"}_{cde}$ 
and the relation (\ref{f-f'}) from theorem \ref{T1}.

(b) $m_{b} = 0$;

In this case we obtain the identity:
\be
f'_{cab} f^{"}_{cde} + f'_{cdb} f^{"}_{cae} +f'_{ceb} f^{"}_{cad} = 0.
\label{ab0}
\ee

If one of the masses
$m_{a}, m_{d}, m_{e}$
is non-zero, we can take, say
$m_{a} \not= 0$
without losing generality and we obtain again an identity. So, the relation
(\ref{ab0}) gives something non-trivial only in the case
$m_{a} = m_{d} = m_{e} = 0$;
in this case the condition (\ref{h3-0}) from the statement is obtain.
\item
We now consider the integrated Wick monomial
$
\int_{\R^{4}} g_{\epsilon}^{2}:u_{a} \Phi_{b} \Phi_{d} \Phi_{e} \Phi_{f}:
$
and we will obtain an expression for
$\tilde{g}^{(8)}_{abdef}$.
As before, we have two cases:

(a) $m_{a} \not= 0$.

In this case we get an expression for $\tilde{g}^{(8)}$:
\be
\tilde{g}^{(8)}_{abdef} \equiv -{i \over 5m_{a}} 
\left( f'_{cba} g_{cdef} +f'_{cda} g_{cbef} +f'_{cea} g_{cbdf} +
f'_{cfa} g_{cbde} \right).
\label{g8}
\ee

One can see easily that if two of the masses
$m_{b}, m_{d}, m_{e}, m_{f}$
are non-zero, then we have
$\tilde{g}^{(8)}_{abdef} = 0$.
If only one of these masses, say 
$m_{b}$
is non-zero, one should have symmetry in the indices $a$ and $b$. This is
indeed a consequence of relation (\ref{f-f'}).

(b) $m_{a} = 0$

In this case we get the following condition of absence of anomaly:
\be
f'_{cba} g_{cdef} +f'_{cda} g_{cbef} +f'_{cea} g_{cbdf} + f'_{cfa} g_{cbde} 
= 0. 
\ee

One can see easily that if one of the masses
$m_{b}, m_{d}, m_{e}, m_{f}$
is non-zero, then we have an identity. If all these masses are zero we get item
(f) from the statement.
\item
At last, we consider now the coefficient of the Wick monomial
$
\int_{\R^{4}} dx g^{2}_{\epsilon} :u_{a} \Phi_{b} \Phi_{c}:
$
we get
\be
3i \tilde{h}^{(3)}_{abc} m _{a} + 
2 \left(f'_{dba} h'_{cd} + f'_{dca} h'_{bd} \right) = 0.
\label{h3}
\ee

For
$m_{a} = 0$
we obtain the condition (\ref{h3-2}) from the statement. We also get
\be
\tilde{h}^{(3)}_{abc} = {2i \over 3m_{a}}
\left(f'_{dba} h^{(3)}_{cd} + f'_{dca} h^{(3)}_{bd} \right), \quad
{\rm for} \quad m_{a} \not= 0.
\ee
\end{itemize}
We have obtained all the relations from the statement and it is clear that we
have used completely the equation (\ref{gi-2}).
$\qed$

Now we can extract some group-theoretical informations from this theorem.
\begin{cor}
(a) The expressions
$f_{abc}$
are the structure constants of a Lie algebra $\mathfrak{g}$.

(b) The structure constants
$f_{abc}$
corresponding to
$m_{a} = m_{b} = m_{c} = 0$
generate a Lie subalgebra
$\mathfrak{g}_{0} \subset \mathfrak{g}$.

(c) The 
$r \times r$
(antisymmetric) matrices
$T_{a}, \quad a = 1,\dots,r$
defined according to
\be
(T_{a})_{bc} \equiv - f'_{bca}, \quad \forall a,b,c = 1,\dots,r.
\ee
are an $r$-dimensional representation of the Lie algebra
$\mathfrak{g}$.
\end{cor}

The first assertion follows from (\ref{anti-f}) and (\ref{Jacobi}), the second
from (\ref{mass-f}) and the third from (\ref{f-f'}).

\begin{rem}
The representation
$T_{a}$
exhibited above is nothing else but the representation of the gauge group $G$
into which the Higgs fields live.
\end{rem}

\begin{rem}
Much of the effort from the Appendix of \cite{ASD3} is nothing else but the
painful verification that the standard model fulfils the conditions
(\ref{f-f'}) and (\ref{repr-f'}) and that all other equations are identically
verified. The advantage of our approach consists in exhibiting very clearly
where the computational difficulties are hidden.
\end{rem}

To verify the condition (\ref{repr-f'}) in specific models it is convenient to
detail this relation. We have by an elementary analysis
\begin{cor}
The relation (\ref{repr-f'}) is equivalent to the following set of relations:
\be
\sum_{m_{c} \not= 0} \left(f_{abc} g_{dec} + f_{ebc} f_{dac} \right) = 0,
\quad {\rm for} \quad m_{b} = m_{d} = 0, \quad m_{a} \not= 0, 
\quad m_{e} \not= 0;
\label{15}
\ee
\begin{eqnarray}
2 \sum_{m_{c} \not= 0} f_{abc} g_{dec} m_{c} m_{e} 
+ \sum_{m_{c} \not= 0} {1 \over m_{c}}
\left[ f_{ceb} g_{dca} m_{a} \left( m_{c}^{2} + m_{e}^{2} - m_{b}^{2}\right)
- (a \leftrightarrow b) \right] 
\nonumber \\
+ \sum_{m_{c} = 0} 
\left[ f'_{dca} g_{ceb} m_{b} - (a \leftrightarrow b) \right] = 0, 
\quad {\rm for} \quad m_{d} = 0, \quad m_{e} \not= 0,\quad m_{a} \not= 0, 
\quad m_{b} \not= 0,;
\label{16}
\end{eqnarray}
\begin{eqnarray}
2 \sum_{m_{c} \not= 0} \left( m_{d}^{2} + m_{e}^{2} - m_{c}^{2}\right) 
f_{abc} f_{dec} - \sum_{m_{c} \not= 0} {1 \over m_{c}^{2}}
[ f_{cda} f_{ceb} \left( m_{c}^{2} + m_{d}^{2} - m_{a}^{2}\right)
\left( m_{c}^{2} + m_{e}^{2} - m_{b}^{2}\right)
\nonumber \\
- (a \leftrightarrow b) ] 
- 4 m_{a}m_{b} m_{d} m_{e} \sum_{m_{c} = 0} 
\left[ g_{dca} g_{ceb} - (a \leftrightarrow b) \right] = 0, 
\quad {\rm for} \quad m_{d} \not= 0, \quad m_{e} \not= 0; 
\label{17}
\end{eqnarray}
\begin{eqnarray}
\sum_{m_{c} \not= 0} f_{abc} f'_{dec} 
- m_{a}m_{b} \sum_{m_{c} \not= 0} 
\left( g_{dca} g_{ceb} - (a \leftrightarrow b) \right) 
- \sum_{m_{c} = 0} \left(f'_{cda} f'_{ceb} - (a \leftrightarrow b) \right) = 0
\nonumber \\
{\rm for} \quad m_{d} = m_{e} = 0.
\label{18}
\end{eqnarray}
\end{cor}

{\bf Proof:}
One considers the separately  distinct cases of (\ref{repr-f'}) namely when
$m_{d}$
and
$m_{e}$
are both equal to $0$, both non-null or only one of them is equal to $0$ and
obtains respectively (\ref{18}), (\ref{17}) and (\ref{15})+(\ref{16}) if one
substitutes the explicit expressions (\ref{f'1}) and (\ref{f'2}) for
$f'_{abc}$.
$\qed$

Now we have as in \cite{Gr1} Corollary 4.8:
\begin{cor}
Suppose that the constants 
$f_{abc},~ f_{abc},~ h_{ab}$
and
$h'_{ab}$
verify the conditions from the statements of theorems \ref{T1} and \ref{T2}.
Then, the general expression for the chronological product
$T_{2}$
is given by the sum of the particular solution:
\begin{eqnarray}
T_{2}^{c}(x,y) =  :T_{1}(x) T_{1}(y): + T_{2}^{0}(x,y) + T_{2}^{h}(x,y) + 
i\delta (x-y) 
\nonumber \\
\times [ {1\over 4} f_{cab} f_{cde} 
:A_{a\nu}(x) A_{b\nu}(x) A_{d}^{\mu}(x) A_{e}^{\nu}(x): 
- f'_{cda} f'_{ceb} 
:A_{a\nu}(x) A_{b}^{\nu}(x) \Phi_{d}(x) \Phi_{e}(x): 
\nonumber \\
+ \sum_{m_{a}, m_{b}, m_{d}, m_{e} \not= 0} {3 \over 2m_{b}} 
f'_{cab} f^{"}_{cde} 
:\Phi_{a}(x) \Phi_{b}(x) \Phi_{d}(x) \Phi_{e}(x):
\nonumber \\
- \sum_{m_{a} \not= 0} {4\over 5m_{a}} f'_{cba} g_{cdef}
:\Phi_{a}(x) \Phi_{b}(x) \Phi_{d}(x) \Phi_{e}(x) \Phi_{f}(x):
\nonumber \\
- \sum_{m_{a} \not= 0} {2 \over 3m_{a}} f'_{dba} h'_{cd} 
:\Phi_{a}(x) \Phi_{b}(x) \Phi_{c}(x): + L(x)];
\label{T-2}
\end{eqnarray}
here the expressions 
$T_{2}^{0}(x,y)$
and
$T_{2}^{h}(x,y)$
have been defined previously according to the formul\ae~ (\ref{T2-generic}) 
and (\ref{T2-h})) and the Wick monomial 
$L(x)$
is an finite normalisation of the type (\ref{YM-1}). In particular, the theory
is renormalisable up to order two.  The condition of unitarity can be satisfied
if and only if
$
L(x)^{\dagger} = L(x)
$
\end{cor}

We only note that the expression of the finite normalisation follow from the
expressions (\ref{g1}), (\ref{g5}), (\ref{g6}), (\ref{g7}), (\ref{g8}), and 
(\ref{h3}) of
$\tilde{g}^{(1)}_{abcd}$,
$\tilde{g}^{(5)}_{abcd} - \tilde{g}^{(8)}_{abcd}$
and
$\tilde{h}^{(3)}_{abc}$.

\begin{rem}
It was noticed in \cite{DHKS1} that the expression 
$T_{11}$
from theorem \ref{T1} and the first finite normalisation from the preceding
formula reconstruct the usual Yang-Mills Lagrangian. A similar remark is in
order in this context, namely the expression
$T_{12}$
from theorem \ref{T1} and the second finite normalisation from the preceding
formula reconstruct the usual kinematic part of the Higgs Lagrangian (see for
instance \cite{We}.) The next terms are part of the Higgs potential.
\end{rem}
\newpage

\subsection{The Standard Model}

It was clear from the preceding sections that in order to specify a certain
concrete model of heavy spin-one Bosons it is not sufficient to specify the
gauge group $G$ from theorem \ref{T1} but also to fix a basis in the Lie
algebra 
$Lie(G)$.
This is a consequence of the fact that the assignment of the masses
$m_{a},~m_{b}$,
etc. is connected with a specific basis and if we choose another basis we will
obtain fields which do not create particles of fixed mass.

For the case of the standard model it means that we have to specify the group,
which in this case is
$SU(2) \times U(1)$
{\it and}
the basis through the {\it Weinberg angle}. Explicitly, let us take in the Lie
algebra of
$SU(2) \times U(1)$
the standard basis 
$X_{a}, \quad a = 0,1,2,3$
with the commutation relations
\be
[ X_{a}, X_{b}] = \epsilon_{abc} X_{c}, \quad a, b = 1,2,3, \quad
[ X_{0}, X_{a}] = 0, \quad a = 1,2,3.
\ee

We consider another basis 
$Y_{a}, \quad a = 0,1,2,3$
defined by
\be
Y_{a} = g~X_{a}, \quad a = 1,2, \quad
Y_{3} = - g~cos \theta~ X_{3} + g'~sin~\theta X_{0}, \quad
Y_{0} = - g~sin \theta~ X_{3} - g'~cos~\theta X_{0}.
\ee

By definition, the angle $\theta$, determined by the condition
$cos~\theta > 0$
is called the {\it Weinberg angle}. The constants $g$ and $g'$ are real with 
$g > 0$.
Then one can show that the new commutation rules produce the following
structure constants \cite{ASD3}:
\be
f_{210} = g~sin~\theta, \quad f_{321} = g~cos~\theta, \quad 
f_{310} = 0, \quad f_{320} = 0
\label{standard-const}
\ee
and the rest of the constants are determined by antisymmetry. By definition,
the {\it standard model} corresponds to this choice of constants {\it and} to
the following assignment of masses:
\be
m_{0} = 0, \quad m_{a} \not= 0, \quad a = 1,2,3.
\ee

We say that the particles created by
$A^{\mu}_{0}$
are {\it photons} and the particles created by
$A^{\mu}_{a}, \quad a = 1,2,3$
are {\it heavy Bosons} (more precisely, for 
$a = 1, 2$
we have the {\it W-Bosons} and for
$a = 3$
the {\it Z-Boson}).

We will derive bellow, directly from our general analysis, that the standard
model is compatible with all restrictions outlined in the previous analysis and
we will see that the only free parameters are, essentially
$m^{H}_{0}$,
$f^{"}_{000}$
and
$g_{0000}$. 

\begin{thm}
In the standard model, the following relations are true:

(a) the masses of the heavy Bosons are constrained by:
\be
m_{1} = m_{2} = m_{3} cos~\theta;
\ee

(b) the constants
$f'_{abc}$
are completely determined by the antisymmetry property (\ref{anti-f'}) and:
\be
f'_{011} = f'_{022} = {\epsilon~g\over 2}, \quad
f'_{033} = {\epsilon~g \over 2cos~\theta}, \quad
f'_{210} = g~sin~\theta, \quad
f'_{321} = - f'_{312} = {g \over 2}, \quad
f'_{123} = - g~{cos~2\theta \over 2cos~\theta},
\ee
the rest of them being zero. Here 
$\epsilon$
can take the values $+$ or $-$.

(c) the constants
$f^{"}_{abc}$
are (partially) determined by:
\begin{eqnarray}
f^{"}_{abc} = 0 \quad (a,b,c = 1,2,3), \quad
f^{"}_{001} = f^{"}_{002} = f^{"}_{003} = 
f^{"}_{012} = f^{"}_{023} = f^{"}_{031} = 0, 
\nonumber \\
f^{"}_{011} = f^{"}_{022} = f^{"}_{033} = 
{\epsilon g\over 12 m_{1}} (m_{0}^{H})^{2}. 
\end{eqnarray}

(d) The constants 
$h_{ab}$
and 
$h'_{ab}$
are determined by
\be
h_{ab} = 0, \quad a \not= b, \quad
h_{11} = h_{22} = h_{33} = 2 h'_{00}.
\ee
\end{thm}

{\bf Proof:}
(i) We first consider the consistency relation (\ref{mass-f}) and immediately
get that
$m_{1} = m_{2}$.
The consistency condition (\ref{mass-f'}) is trivial because from the
antisymmetry property (\ref{anti-f'}) we have:
\be
f'_{00a} = 0, \quad a = 0,1,2,3.
\label{f00}
\ee

(ii) We investigate now the consistency condition (\ref{repr-f'}). It is
convenient to use it in the detailed form (\ref{15}) - (\ref{18}). We mention
briefly the result of elementary computations. From (\ref{15}) we obtain
equivalently that 
\be
g_{0ab} = 0, \quad a,b = 1,2,3, \quad a \not= b, \quad 
g_{011} = g_{022}.
\ee

From (\ref{16}) only the case
$a = 1,~b = 2$
gives something non-trivial, namely
\be
g_{033} = g_{011} = g_{022}.
\ee

Next, we consider the relation (\ref{17}). From the case
$d = 1,~e = 2,~a = 1,~b = 2$
we get
\be
m_{1}^{2} g_{011}^{2} = g^{2}~
\left( 1 - {3 m_{3}^{2} \over 4 m_{1}^{2}} cos^{2}~\theta\right)
\ee
and from the case
$d = 1,~e = 2,~a = 2,~b = 3$
\be
g_{022} g_{033} = g^{2}~{m_{3}^{2} \over 4 m_{1}^{4}} cos^{2}~\theta;
\ee
all other cases give identities. Observe now that the last two relations are
consistent {\it iff} we have
\be
m_{1} = m_{3} cos~\theta.
\ee

Finally, the relation (\ref{18}) is trivial.  From the preceding relations, we
can reconstruct all the constants
$f'_{abc}$
as given in the statement.

(iii) From (\ref{f"}) we get the expressions of the constants
$f^{"}_{abc}$
from the statement. We also observe that the relation (\ref{h3-0}) is
identically verified.

(iv) The relation (\ref{h3-2}) gives
\be
h_{ab} = 0, \quad a \not= b, \quad
h_{11} = h_{22} = h_{33} = 2 h'_{00}.
\ee

(v) Finally, the conditions (\ref{f'g}) give no restrictions.  We have obtained
all the relations from the statement.  
$\qed$

\begin{rem}
In the standard model one disregards the terms 
$T_{15}$ 
and 
$T_{16}$ 
from theorem \ref{T1} and it follows that the expression
$T_{2}^{h}$
can be put to zero. One can also show \cite{AS} that it is possible to fix
$\epsilon = +$.
We note that we are left only with the Yang-Mills interaction of the usual
form. However, now we can do all the computations completely rigorously after
we have the convenient splitting of the distributions involved in the analysis.

The choice
$g = {e\over sin~\theta}, \quad g' = - {e\over cos~\theta}$
can be obtained if one includes interaction with matter and requires that the
interaction of the electron Dirac field with the electromagnetic potential has
the usual form.
\end{rem}

We note in the end two facts. First, we note that because of the equality 
$m_{1} = m_{2}$
there exists a global symmetry of the theory generated by the {\it electric
charge} which commutes with the $S$-matrix.

Next, suppose we admit that the photon has a
small non-zero mass 
$m_{0} \not= 0$ 
and we try to interpret the adiabatic limit as the the process
$lim_{m_{0} \searrow 0} lim_{\epsilon \searrow 0}$.
One can easily prove that this is not possible. Indeed, if all the masses
$m_{a}, \quad a = 0,\dots,3$
are non-zero, that the expressions
$f'_{abc}$
are given by the expressions (\ref{f'1}) for all values of the indices. One can
plug this expression into the relation (\ref{repr-f'}). If one considers, for
instance, the cases
$a = d = 0, \quad b = e = 1$
and obtain the following relation
\be
m_{0}^{4} + 2 (m_{2}^{2} - m_{1}^{2}) m_{0}^{2} + 
(m_{1}^{2} - m_{2}^{2}) (m_{1}^{2} + 3 m_{2}^{2}) = 0.
\label{0011}
\ee

In the case
$a = d = 0, \quad b = e = 2$
one obtains the preceding relation with
$m_{1} \leftrightarrow m_{2}$.
If we subtract the two relations, then we get
\be
(m_{1}^{2} - m_{2}^{2}) (m_{0}^{2} - m_{1}^{2} - m_{2}^{2}) = 0.
\ee

We have two cases: if
$m_{1} = m_{2}$
then from (\ref{0011}) we obtain
$m_{0} = 0$;
if
$m_{0}^{2} = m_{1}^{2} + m_{2}^{2}$
then again (\ref{0011}) gives
$m_{1} m_{2} = 0.$
So, we obtain that at least one of the masses
$m_{a}, \quad a = 0,1,2$
must be null, which contradicts the hypothesis that all masses are non-zero.

\newpage
\section{Conclusions}

We have analysed in full generality the possibilities of coupling non-trivially
heavy Bosons of spin one up to order two of the perturbation theory. In
particular we can reobtain in a rather elementary way the standard model
(without leptons). In a subsequent publication \cite{Gr2} we investigate, in
our more general framework, the case when the leptons are included. In
particular it is expected according to the usual analysis (see also
\cite{ASD3}) that,  going to the third order of the perturbation theory, we
will find out new restrictions on the parameters
$f^{"}_{000}$ 
and
$g_{0000}$
and also some restrictions on the Fermion sector, namely the cancellation of
some anomaly (of Adler-Bell-Jackiw-type).

Another extremely interesting problem is to investigate the class of Lie groups
for which there exists a non-trivial solution to our problem. Indeed, it is not
obvious that any Lie group of the type described in the statement of theorem
\ref{T1} admits a representation of dimension equal to the dimension of the
group, realised by antisymmetric matrices and verifying the mass relation
(\ref{f-f'}). In the absence of a general solution, one should test the
existence of a non-trivial solution of the perturbation series, at least, for a
simple Lie group like 
$SU(5)$ 
because such groups are characteristic for grand unified theories.

Finally, one should find explicit expressions for the distributions of the type
$D^{F}_{m_{a}m_{b}\cdots}$
and perform rigorous computations for various cross sections of the standard
model. In this way one could check if some differences with respect to the
usual computational approaches to the standard model appear or, more probably,
prove that one obtains the same results. 

\end{document}